\documentclass[a4paper,fleqn,usenatbib]{mnras}

\usepackage{newtxtext,newtxmath}
\usepackage{gensymb}

\usepackage[T1]{fontenc}
\usepackage{ae,aecompl}

\usepackage{siunitx}

\usepackage{graphicx}	
\usepackage{amsmath}	
\usepackage{nccmath}
\usepackage{longtable}
\usepackage{arydshln}
\usepackage{booktabs}
\usepackage{tabularx}
\usepackage{supertabular}
\usepackage{enumitem}
\setlist{nosep}

\title[Rotators within the cosmic web]{The SAMI Galaxy Survey: large-scale environment affects galaxy spin amplitudes and the formation of slow rotators}

\author[S.\ Barsanti et al.]{Stefania Barsanti,$^{1,2,3}$\thanks{E-mail: stefania.barsanti@sydney.edu.au}
Scott M. Croom,$^{1,3}$
Matthew Colless,$^{2,3}$
Joss Bland-Hawthorn,$^{1,3}$
\newauthor
Sarah Brough,$^{4,3}$ 
Julia J. Bryant,$^{1,3,5}$ 
Nuria Lorente,$^{6}$ 
Sree Oh,$^{7,2,3}$
Giulia Santucci,$^{8,3}$
\newauthor
Sarah Sweet,$^{9,3}$ 
Jesse van de Sande,$^{4,2,3}$
Charlotte Welker$^{10,11,12}$  
\vspace{0.4cm}
\\
$^{1}$Sydney Institute for Astronomy (SIfA), School of Physics, The University of Sydney, NSW 2006, Australia\\
$^{2}$Research School of Astronomy and Astrophysics, Australian National University, Canberra, ACT 2611, Australia\\
$^{3}$ARC Centre of Excellence for All Sky Astrophysics in 3 Dimensions (ASTRO 3D), Australia\\
$^{4}$School of Physics, University of New South Wales, NSW 2052, Australia\\
$^{5}$Australis-USydney, School of Physics, University of Sydney, NSW 2006, Australia\\
$^{6}$AAO-Macquarie, Macquarie University, NSW 2109, Australia\\
$^{7}$Department of Astronomy, Yonsei University, Seoul 03722, Republic of Korea\\
$^{8}$International Centre for Radio Astronomy Research, University of Western Australia, 35 Stirling Highway, Crawley, WA 6009, Australia\\
$^{9}${School of Mathematics and Physics, University of Queensland, Brisbane, QLD 4072, Australia}\\
$^{10}$New York City College of Technology, City University of New York, 300 Jay Street, Brooklyn, NY, USA\\
$^{11}$Centre for Computational Physics, Flatiron Institute, 160 5th avenue, New York, NY, USA\\ $^{12}$Department of Physics \& Astronomy, Johns Hopkins University, Baltimore, MD 21218, USA\
}

\date{Accepted 2025 March 10. Received 2025 March 06; in original form 2024 December 20. }

\pubyear{2024}

\begin{document}
\label{firstpage}
\pagerange{\pageref{firstpage}--\pageref{lastpage}}
\maketitle

\begin{abstract}
We explore the impact of the large-scale 3D density field, as defined by deep, wide-field galaxy surveys, on stellar spin ($\lambda_{\rm R_e}$) and the distributions of fast and slow rotators. We use the GAMA spectroscopic redshift survey to reconstruct the cosmic web and obtain spatially-resolved stellar kinematics from the SAMI Galaxy Survey. Among various local and large-scale environment metrics, the distance to the closest filament ($D_{\rm fil}$) correlates most significantly with $\lambda_{\rm R_e}$, but it is secondary to the more dominant roles played by stellar age and mass. Fast rotators tend to have increasing $\lambda_{\rm R_e}$ going from nodes to filaments to voids, independently of mass. Slow rotators and mass-matched fast rotators are found to have significantly different distributions of large-scale environment metrics but consistent distributions of local environment metrics. About 95\% of slow rotators have $D_{\rm fil}\leq2$\,Mpc, while covering broader ranges (similar to fast rotators) in distance to nodes and voids, local galaxy density, halo mass, and position with respect to the halo. At fixed mass, the fraction of slow rotators, $f_{\rm SR}$, increases for smaller $D_{\rm fil}$, especially for massive galaxies. While controlling for age or mass, only galaxies very close to filaments and nodes show a significant impact of local environment on $f_{\rm SR}$. Our results demonstrate that the cosmic web leaves an imprint on galactic spin amplitudes, and that pre-processing by mergers occurring within filaments is likely to be an important physical mechanism for the formation of slow rotators before they reach nodes.
\end{abstract}



\section{Introduction}
\label{Introduction}

Understanding how the angular momentum of galaxies is acquired and lost is crucial to comprehending the processes of galaxy formation and evolution. According to tidal torque theory, the spin of a galaxy is generated through torques acting on the collapsing proto-halo, which acquires angular momentum from gravitational perturbations in the tidal field \citep{Hoyle1951,Peebles1969,Doroshkevich1970,White1984,Porciani2002,Schafer2009}. The alignment of the galaxy's spin axis relative to the orientation of the closest cosmic filament retains a record of the galaxy's formation. 

Galaxy spin--filament alignments have been studied in cosmological hydrodynamical simulations \citep{Dubois2014,Welker2014,Codis2018,Wang2018,Kraljic2020}, which suggest that low-mass galaxies formed via gas-accretion mechanisms tend to have spins aligned parallel to the filament they are embedded within, while high-mass galaxies formed via mergers tend to have spins aligned perpendicular to the filament. Observations confirm such trends  \citep{Tempel2013a,Tempel2013b,Pahwa2016,Hirv2017,Chen2019,BlueBird2020,Tudorache2022}, with the latest studies taking advantage of integral field spectroscopy (IFS) for more precise measurement of galaxies' spin axes from spatially-resolved stellar kinematic maps \citep{Welker2020,Kraljic2021,Barsanti2022,Barsanti2023}. These results tie galaxy angular momentum to the large-scale structure of the Universe, highlighting the impact of the early environment on a galaxy's spin.

One of the most outstanding outcomes of IFS galaxy surveys such as SAURON \citep{Bacon2001}, ATLAS3D \citep{Cappellari2011a}, MaNGA \citep{Bundy2015}, CALIFA \citep{Sanchez2012} and SAMI \citep{Croom2012,Bryant2015}, has been the introduction of a new classification for galaxies based on stellar kinematic properties. Exploiting the joint distribution of stellar spin parameter ($\lambda_{\rm R_e}$, based on the ratio of ordered to random stellar motion within the galaxy's effective radius) and ellipticity ($\epsilon$), galaxies with high $\lambda_{\rm R_e}$ are classified as fast rotators while those below a $\lambda_{\rm R_e}$ versus $\epsilon$ threshold are called slow rotators (\citealp{Cappellari2007, Emsellem2007, Emsellem2011, Cappellari2016, vandeSande2021a}).  In the local volume, about 85\% of the stellar mass is in fast rotators (FRs) characterised by ordered rotation and stellar discs (\citealp{FraserMcKelvie2022}), while about 15\% of the stellar mass is in slow rotators (SRs) showing complex kinematic features and dispersion-dominated stellar motions \citep{Cappellari2016}. 

The role played by the environment in the formation of SRs has been widely investigated. \citet{Cappellari2011} found that the fraction of SRs shows a twofold increase in the central regions of the Virgo galaxy cluster relative to lower-density environments. This kinematic morphology--density relation within clusters or massive groups has been detected by many studies \citep{DEugenio2013, Houghton2013, Scott2014, Fogarty2014}. Environment is expected to be connected to stellar spin, since galaxy--galaxy mergers are predicted to be a common physical mechanism responsible for kinematic transformation from rotation-dominated to dispersion-dominated motions and the formation of SRs \citep{Naab2014,Penoyre2017,Lagos2018}. However, internal galaxy properties, such as stellar mass and age, are discovered to be the primary parameters correlating with stellar spin and the fraction of SRs (\citealp{Brough2017,Veale2017,Greene2018,Vaughan2024,Croom2024}). Nevertheless, \citet{Wang2020},\citet{vandeSande2021b} and \citet{Santucci2023} still find a residual weak dependence of stellar spin on local environment at fixed stellar mass.

Previous studies of the kinematic morphology-density relation focus on local galaxy environments, using metrics such as local galaxy density and halo mass. To clarify and broaden the connection between environment and galaxy angular momentum, we explore here the role played by the large-scale structure of our Universe in setting the amplitude of stellar spin. Taking advantage of the SAMI Galaxy Survey for spatially-resolved stellar kinematics and of the highly-complete GAMA spectroscopic redshift survey to reconstruct the cosmic web, we investigate the distributions of SRs and FRs in relation to nodes, filaments, and voids. Our goal is to better understand the physical scales at which environment plays a role in reducing stellar spin, and so elucidate SR formation pathways.

This paper is structured as follows: Section~\ref{Data and Galaxy Sample} describes our galaxy sample and its features; Section~\ref{Results} presents our results on the kinematic properties of galaxies as a function of other internal properties, local environment metrics, and large-scale environment metrics; Section~\ref{Discussion} compares our findings to previous studies and discuss their physical interpretation; and our conclusions are stated in Section~\ref{Summary and conclusions}. Throughout this work, we assume $\Omega_{m}=0.3$, $\Omega_{\Lambda}=0.7$ and $H_0=70$\,km\,s$^{-1}$\,Mpc$^{-1}$.

\section{Data and Galaxy Sample}
\label{Data and Galaxy Sample}

\subsection{The SAMI Galaxy Survey}
\label{SAMI galaxy survey}

For measurements of stellar kinematics we use the SAMI Galaxy Survey, a spatially-resolved spectroscopic survey of 3068 galaxies with stellar masses $\log(M_{\star}/M_{\odot})$\,=\,8--12 and redshifts $0.004<z\leq0.115$ \citep{Bryant2015,Croom2021}. Most of the SAMI targets belong to the three equatorial fields (G09, G12 and G15) of the Galaxy And Mass Assembly survey \citep[GAMA;][]{Driver2011}. Eight massive clusters were also observed \citep{Owers2017}, but are excluded from our analysis due to our choice of local and large-scale environment metrics. 

The Sydney--AAO Multi-object Integral-field spectrograph (SAMI) was mounted on the 3.9\,m Anglo-Australian Telescope \citep{Croom2012}. The instrument had 13 fused optical fibre bundles (hexabundles), each containing 61 fibres of 1.6\,arcsec diameter so that each integral field unit (IFU) had a 15\,arcsec diameter \citep{Bland2011,Bryant2014}. The SAMI fibres fed the two arms of the AAOmega spectrograph \citep{Sharp2006}. The SAMI Galaxy Survey used the 580V grating in the blue arm, giving a resolving power of $R$=1812 and wavelength coverage of 3700--5700\,\AA, and the 1000R grating in the red arm, giving a resolving power of $R$=4263 over the range 6300--7400\,\AA. The median full-width-at-half-maximum values for each arm are FWHM$_{\rm blue}$=2.65\,\AA\ and FWHM$_{\rm red}$=1.61\,\AA\ \citep{vandeSande2017}. The SAMI data-cubes are characterised by a grid of 0.5$\times$0.5\,arcsec spaxels, with blue and red spectra having pixels corresponding to 1.05\,\AA\ and 0.60\,\AA\ respectively.

\subsection{The GAMA survey}
\label{GAMA galaxy survey}
For tracing the cosmic web we use the Galaxy And Mass Assembly survey (GAMA; \citealp{Driver2011,Hopkins2013,Baldry2018,Driver2022}), a spectroscopic and photometric survey of $\sim$300,000 galaxies with $r \le 19.8$\,mag that covers $\sim$286\,deg$^{2}$ in 5 regions called G02, G09, G12, G15 and G23. The redshift range of the GAMA sample is $0<z<0.5$, with a median of $z\sim0.25$. Most of the spectroscopic data were obtained using the AAOmega multi-object spectrograph at the Anglo-Australian Telescope, although GAMA also incorporates previous spectroscopic surveys such as SDSS \citep{York2000}, 2dFGRS \citep{Colless2001,Colless2003}, WiggleZ \citep{Drinkwater2010} and the Millennium Galaxy Catalogue \citep{Driver2005}. 

The GAMA survey’s deep and highly complete spectroscopic redshift data, combined with its wide area, high spatial resolution, and broad wavelength coverage make it an ideal galaxy sample for mapping the cosmic web. In particular, the high spectroscopic completeness of GAMA (98.5\% in the equatorial regions; \citealp{Liske2015}) compared to the SDSS survey ($\sim$80\%; \citealp{Driver2022}) and the 2 magnitudes deeper of GAMA versus SDSS, allow us to reconstruct the filamentary structures in more detail on smaller physical scales.

\subsection{Galaxy properties}
\label{Galaxy properties}

\subsubsection{Kinematic Morphology Classification}
\label{Kinematic Morphology Classification}

To determine the stellar spin of a galaxy ($\lambda_{\rm R_e}$), we take advantage of spatially-resolved stellar kinematics (see \citealp{vandeSande2017} for a complete description). Briefly, the line-of-sight velocity distributions are obtained from the Penalised Pixel-Fitting software (pPXF; \citealp{Cappellari2004,Cappellari2017}), where the red spectrum is smoothed with a Gaussian kernel to match the spectral resolution of the blue spectrum, then the combined blue and red spectrum is re-binned on a grid of uniform velocity spacing. For each bin, pPXF is run in a multi-step process on each galaxy spaxel to estimate the noise from the fit residual, remove emission lines, and extract velocity and velocity dispersion. The best-fit templates are constructed from the MILES library of stellar spectra \citep{SanchezBlazquez2006,FalconBarroso2011}. 

The observed stellar spin is related to the ratio of ordered versus random motions within the galaxy's effective radius \citep{Emsellem2007,vandeSande2017} by 
\begin{equation}
\lambda_{\rm R_e}= \frac{\sum_i F_i R_i |V_i|}{\sum_i F_i R_i \sqrt{V_i^2+\sigma_i^2}}
\end{equation}
where $F_i$, $R_i$, $V_i$, and $\sigma_i$ are the continuum flux, radius, mean velocity, and velocity dispersion of each spaxel. Intrinsic stellar spins, measured assuming galaxies are observed edge-on as rotating oblate axisymmetric spheroids, are available for the SAMI Galaxy Survey from \citet{vandeSande2021a}. These intrinsic spin measurements include a seeing correction from \citet{Harborne2020} and an aperture correction. 

The galaxies' semi-major axis effective radii ($R_e$) and ellipticities within $R_e$ ($\epsilon_e$) are measured using Multi-Gaussian Expansion fits (MGE; \citealp{Emsellem1994,Cappellari2002}). For the SAMI Galaxy Survey, the MGE technique is applied to $r$-band SDSS images \citep{York2000}; a detailed presentation of the MGE fits can be found in \cite{DEugenio2021}.

Finally, galaxies are assigned a kinematic morphological classification (i.e.\ fast rotator versus slow rotator) according to their position in the $\lambda_{\rm R_e}$ versus $\epsilon_e$ plot. We use the threshold defined by \citet{vandeSande2021a}, but the conclusions of this work do not change if we use the threshold defined by \citet{Cappellari2016} instead.  

\subsubsection{Stellar mass, age, and star formation rate}
\label{Stellar mass and age}

Galaxy stellar masses ($M_\star$) are measured from K-corrected $g$$-$$i$ colours and $i$-band magnitudes for the SAMI Galaxy Survey \citep{Bryant2015}. 

For stellar population ages, we use full spectral fitting measurements from \citet{Vaughan2022}. Briefly, for each galaxy, an overall age is obtained from fitting an aperture spectrum obtained by summing the spectra from all spaxels within a $1\,R_e$ ellipse \citep{Scott2018}. The procedure makes use of pPXF and the MILES simple stellar population models \citep{Vazdekis2015}. Gas emission lines are fit simultaneously with the stellar population models and a $10^{\rm th}$-order multiplicative polynomial is used. Light-weighted (i.e.\ luminosity-weighted) and mass-weighted ages are both obtained. In this work, we make use of light-weighted ages, since \citet{Croom2024} show that $\lambda_{\rm R_e}$ is more strongly correlated with light-weighted ages than with mass-weighted ages, stellar mass, or local environment.

The star-formation rate (SFR) is estimated using dust-corrected H$\alpha$ flux, assuming an intrinsic Balmer decrement H$\alpha$/H$\beta$\,=\,2.86 and the \citet{Cardelli1989} extinction law. Only star-forming spaxels, identified using Baldwin, Phillips \& Terlevich (BPT) emission-line diagnostics \citep{Baldwin1981,Veilleux1987,Medling2018}, are taken into account. We use integrated SFR values within the galaxy's semi-major effective radius $R_{\rm e}$. These measurements tend to underestimate the total SFR, since they are limited to 1\,$R_{\rm e}$ aperture, and to be accurate for star-forming dominant galaxies and less reliable for passive and AGN galaxies. However, the same results are found by using photometric SFRs estimated for SAMI galaxies by \citet{Ristea2022}, meaning our results are not biased by emission activity.

\subsection{Local environment metrics}
\label{Local environment metrics}

To investigate the local environment we use three different metrics, following \citet{Croom2024}: local galaxy density, halo mass, and environmental classification. 

Local galaxy density is measured as the fifth-nearest neighbour surface density $\Sigma_5$ \citep{Brough2013}. We use the GAMA galaxy group catalogue built with a friend-of-friends algorithm \citep{Robotham2011}, where halo mass is estimated using group velocity dispersion and projected distance. Finally, each galaxy is classified either as a central group galaxy or as a satellite (if it is a group member), or as isolated (if it does not belong to any GAMA group). We call this latter environmental classification `local type'.

\subsection{Large-scale environment metrics}
\label{Large-scale environment metrics}

We use the cosmic web reconstruction for the GAMA G12, G13 and G15 regions from \citet{Barsanti2022}. The reconstruction takes advantage of the Discrete Persistent Structure Extractor public code ({\sc DisPerSE}; \citealp{Sousbie2011a,Sousbie2011b}). {\sc DisPerSE} is a parameter-free and scale-free topologically-motivated algorithm, based on discrete Morse and persistence theories. Voids, walls, and filaments are identified as distinct regions of the cosmic web by taking advantage of galaxies' sky positions and spectroscopic redshifts. They are defined as distinct regions in the geometrical segmentation of space following the gradient of the Morse function. A void is defined as a 3D structure that originates from a given minimum, walls are 2D structures that delimit voids, and filaments are 1D structures that join maxima together.

The identification of cosmic filaments can be affected by the `Fingers of God' effect (FoG; \citealp{Jackson1972}), the elongation of halos in redshift space along the line of sight due to the random motions of galaxies within virialised groups and clusters. These distorted halos can lead to the identification of spurious filaments, although \citet{Barsanti2022} conclude that corrections for the FoG effect do not make significant changes to the cosmic web inferred for the SAMI Galaxy Survey.

\begin{figure}
\includegraphics[width=8cm]{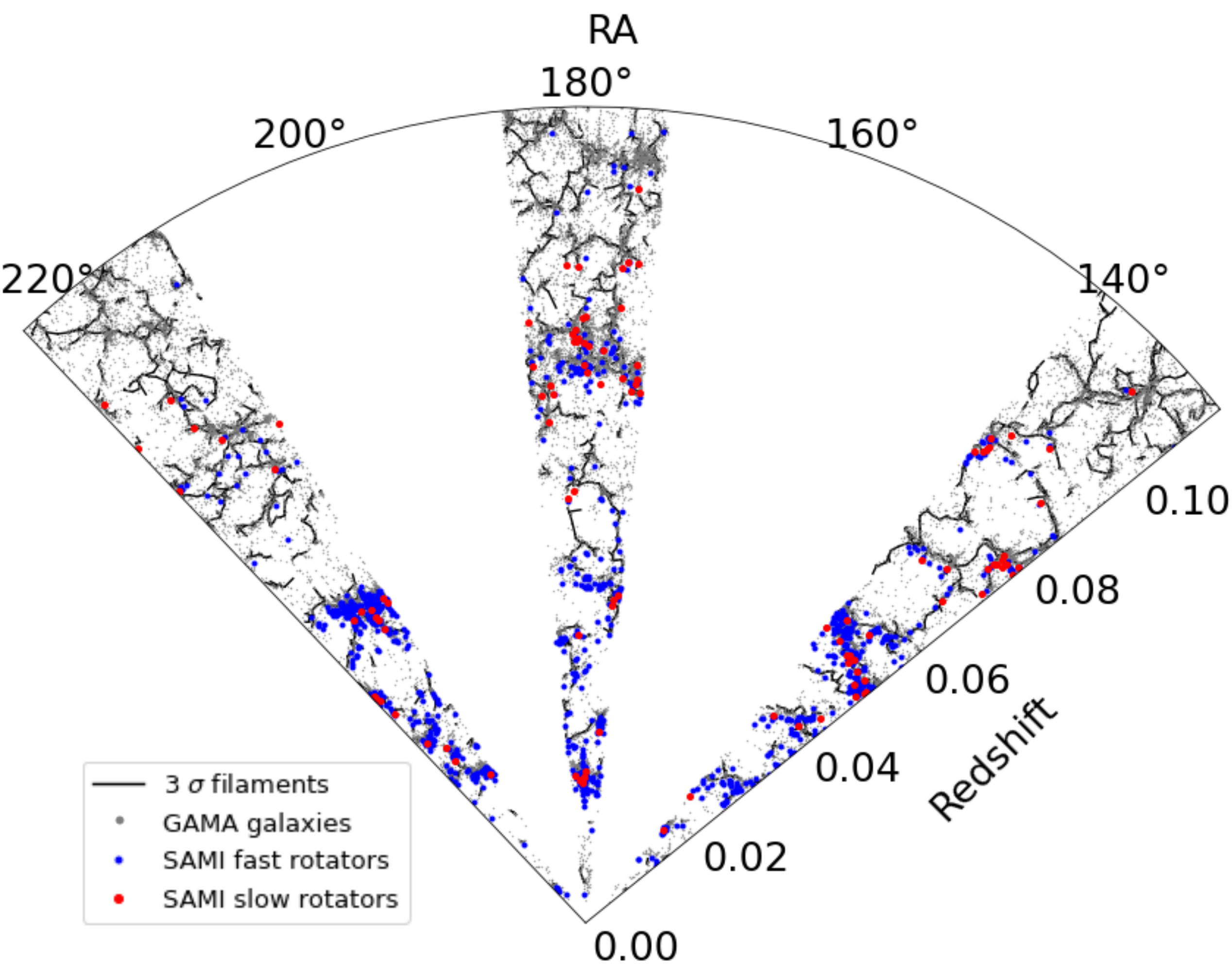}
\caption{Projected network of cosmic filaments (blue lines) for the GAMA G09, G12 and G15 regions. Grey points are the 35882 GAMA galaxies used to reconstruct the spine of the cosmic web, blue points are 976 SAMI fast rotators, and red points are 118 SAMI slow rotators.}
\label{PolarPlotSAMIGAMAFilaments}
\end{figure}

For the galaxy distribution, \citet{Barsanti2022} use 35882 GAMA galaxies with secure redshifts and stellar masses in the SAMI redshift range ($0<z<0.13$) within the G09, G12 and G15 regions. {\sc DisPerSE} is then run with a 3$\sigma$ persistence threshold to identify the most significant structures. The 3D filamentary structure is shown as an interactive plot at \href{https://skfb.ly/o9MXz}{this URL}; Figure~\ref{PolarPlotSAMIGAMAFilaments} shows the projected network of filaments.

\begin{figure}
\centering
\includegraphics[width=6cm]{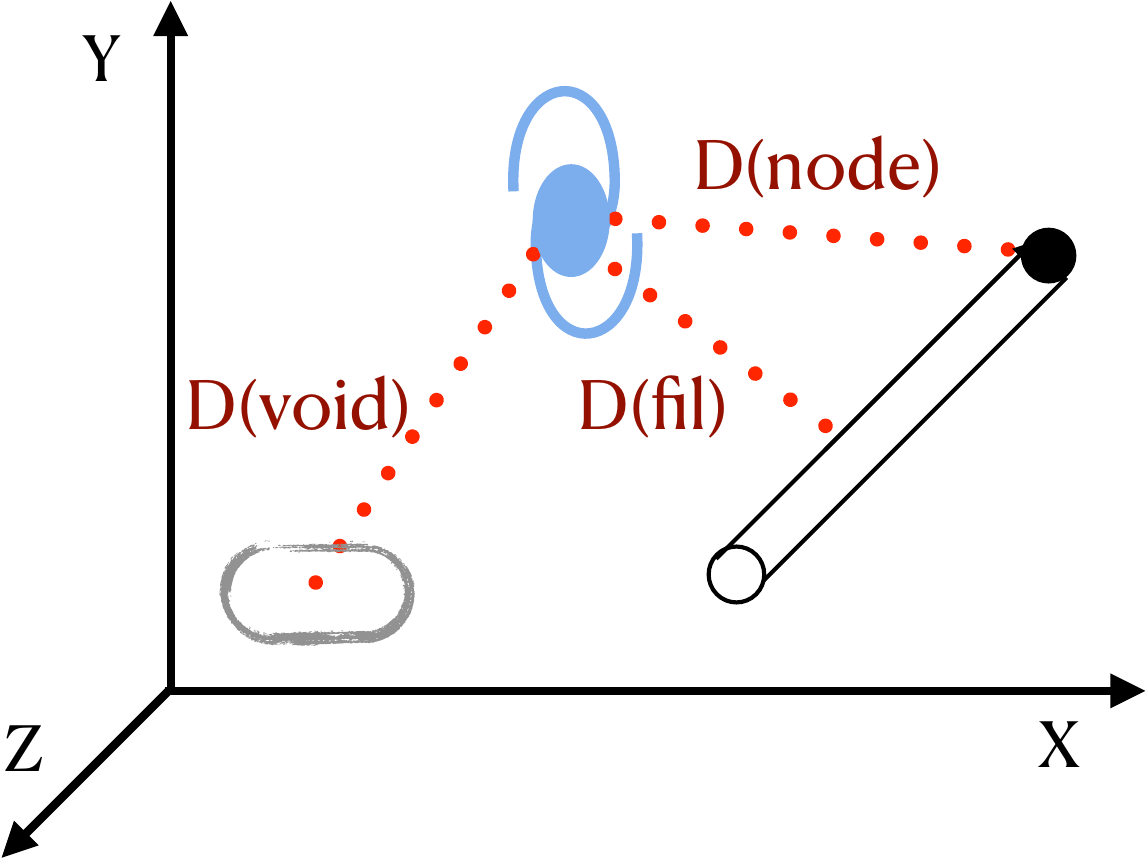}
\caption{Schematic representation of the cosmic web metrics: distance to node ($D_{\rm node}$), distance to filament ($D_{\rm fil}$), and distance to void ($D_{\rm void}$).}
\label{CosmicWebMetrics}
\end{figure}

To understand the distributions of FRs and SRs within the cosmic web and the correlation between $\lambda_{\rm R_e}$ and the large-scale environment, we make use of the following cosmic web metrics: distance to node ($D_{\rm node}$), distance to filament ($D_{\rm fil}$), and distance to void ($D_{\rm void}$). Following \citet{Barsanti2022}, each SAMI galaxy is assigned to the closest cosmic filament using the smallest 3D Euclidean distance. If the projection point is beyond the start or end of the filament segment, we use the distance to the closest node as $D_{\rm fil}$; by construction, $D_{\rm fil}$ is always smaller than $D_{\rm node}$. A schematic representation of these cosmic web metrics is shown in Figure~\ref{CosmicWebMetrics}.

In Appendix~\ref{Local environments within cosmic regions} we investigate the location of the local environments with respect to the structures identified in the cosmic web and show that these measures of small-scale and large-scale environment, while different, are both consistent with the hierarchical formation of structure.

\subsection{Galaxy sample}
\label{Galaxy sample}

Within the GAMA regions there are 1094 SAMI galaxies with $\log(M_{\star}/M_{\odot})\geq 9.5$, the threshold mass for complete stellar kinematic measurements \citep{vandeSande2021b}, and reliable stellar spins. Each of these galaxies is associated with the closest node, filament, and void. This sample contains 976 FRs and 118 SRs, plotted in relation to the network of cosmic filaments in Figure~\ref{PolarPlotSAMIGAMAFilaments}. Of these, 1062 have measured stellar ages and local galaxy densities, with 950 FRs and 105 SRs having estimates for all these properties; these galaxies are classified into 423 centrals, 302 satellites, and 337 isolated. Of these 1062 galaxies, 725 belong to groups with available halo masses, of which 630 are FRs and 95 are SRs. 

\section{Results}
\label{Results}

In order to better understand at which scale(s) environment has a role in influencing stellar spin, we study the connection between $\lambda_{\rm R_e}$ and the 3D cosmic web, and compare to our earlier work on internal galaxy properties and local environment metrics. By investigating the distributions of FRs and SRs across the large-scale structures, we aim to shed light on the physical mechanisms responsible for the formation of SRs.

\subsection{Kinematic segregation within the cosmic web}
\label{Spin amplitude}

Previous observational works focus on investigating the {\it orientation} of a galaxy's spin with respect to cosmic web structures, such as filaments (e.g., \citealp{Tempel2013a,Welker2020,Kraljic2021,Barsanti2022}) and voids \citep{Lee2023}, with the goal of relating preferred orientations to internal galaxy properties. 

Here we seek to understand whether the {\it amplitude} of a galaxy's angular momentum is related to the cosmic web. The stellar spin parameter $\lambda_{\rm R_e}$ is used as a proxy: higher spins indicate more rotation-dominated galaxies. We classify our initial sample of 1094 SAMI galaxies (see Section~\ref{Galaxy sample}) by the cosmic structure they reside within: we identify 230 galaxies with $D_{\rm node}<1$\,Mpc as belonging to halos, 527 galaxies with $D_{\rm node}>1$\,Mpc and $D_{\rm fil}<1.5$\,Mpc as belonging to filaments, and the remaining 337 galaxies as belonging to voids.

\begin{figure*}
\includegraphics[width=1.0\textwidth]{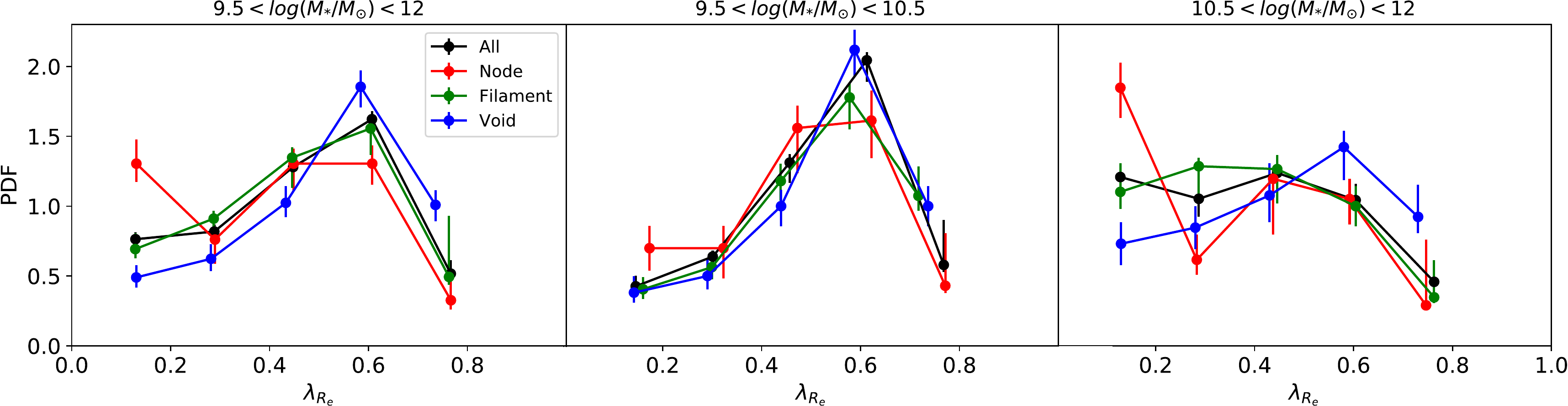}
\caption{Distributions of $\lambda_{\rm R_e}$ for 1094 SAMI galaxies, with the PDF for all galaxies in black, and the PDFs for galaxies belonging to nodes, filaments, and voids in red, green, and blue respectively. The left, middle, and right panels show PDFs for the full mass range ($9.5<\log(M_{\star}/M_{\odot})<12$), low-mass galaxies ($9.5<\log(M_{\star}/M_{\odot})<10.5$), and high-mass galaxies ($10.5<\log(M_{\star}/M_{\odot})<12$), respectively. Error bars are from 1000 bootstrap samples.}
\label{PDFlambda}
\end{figure*}

With these identifications, the left panel of Figure~\ref{PDFlambda} shows the probability distribution function (PDF) of $\lambda_{\rm R_e}$ for galaxies belonging to nodes, filaments, and voids; the middle and right panels show these PDFs for, respectively, 587 low-mass galaxies with $9.5<\log(M_{\star}/M_{\odot})<10.5$ and 513 high-mass galaxies with $10.5<\log(M_{\star}/M_{\odot})<12$. The PDFs are normalised such that the mean value over the bins is unity, and the error bars are estimated from the bootstrap method using 1000 sample realisations. For the whole sample, galaxies with low values of $\lambda_{\rm R_e}$ are most commonly found in nodes and rarely found in voids. High-$\lambda_{\rm R_e}$ galaxies show the opposite tendency, being most commonly found in voids and rarely found in nodes. Low-mass galaxies lack low $\lambda_{\rm R_e}$ values, while those with high-$\lambda_{\rm R_e}$ have a slight preference for being in voids. High-mass galaxies tend to have lower $\lambda_{\rm R_e}$ values, with the lowest $\lambda_{\rm R_e}$ galaxies mostly in nodes and the higher $\lambda_{\rm R_e}$ galaxies in voids.

Table~\ref{PDFlambdaResults} reports results of two-sample Kolmogorov-Smirnov (K-S) tests \citep{Lederman1984} comparing the $\lambda_{\rm R_e}$ distributions of galaxies in nodes, filaments, and voids. For the sample as a whole, galaxies associated with different large-scale structures are likely to be drawn from different stellar-spin populations. For low-mass galaxies only the $\lambda_{\rm R_e}$ distributions within nodes and voids are significantly different ($p_{\rm 2KS}=0.02$). The high-mass subsample again shows three distinct stellar-spin distributions in the different elements of the cosmic web, although with lower statistical significance than for the sample as a whole.

\begin{table*}
\caption{Results from two-sample Kolmogorov-Smirnov tests for the $\lambda_{\rm R_e}$ distributions of galaxies in nodes, filaments, and voids. Columns~1 gives the $M_{\star}$ range; columns~2 and~3 give the subsamples for which $\lambda_{\rm R_e}$ distributions are compared, with the numbers of galaxies in square brackets; the associated $p$-value is listed in column~4; columns~5--7 are the same as columns~2--4, but for fast rotators only. Significant $p$-values (those less than 0.05) are highlighted in bold.}
\centering
\begin{tabular}{cccc|ccc}
\toprule
$M_{\star}$ range &  Sample 1 [N$_{\rm gal}$] & Sample 2 [N$_{\rm gal}$] & $p_{\rm 2KS}$ & Sample 1 [N$_{\rm FR}$] & Sample 2 [N$_{\rm FR}$] & $p_{\rm 2KS}$\\
 \midrule
$ \left[9.5; 12\right]$ &  Nodes [230] & Voids [337] & $\mathbf{10^{-6}}$ & Nodes [184] & Voids [317] & $\mathbf{10^{-3}}$  \\
$\left[9.5; 12\right]$&   Nodes [230]& Filaments [527]  & $\mathbf{10^{-2}}$ & Nodes [184] & Filaments [475] & 0.74 \\
$\left[9.5; 12\right]$  & Filaments [527] & Voids [337] &  $\mathbf{10^{-4}}$ & Filaments [475] & Voids [317] & $\mathbf{10^{-3}}$ \\
\midrule
$\left[9.5; 10.5\right]$ &  Nodes [93]& Voids [210] &  \textbf{0.02}& Nodes [89]& Voids [203] & \textbf{0.03} \\
$\left[9.5; 10.5\right]$&   Nodes [93]& Filaments [284] & 0.41 &   Nodes [89]& Filaments [274] &  0.45\\
$\left[9.5; 10.5\right]$ &  Filaments [284] & Voids [210] & 0.16 &  Filaments [274] & Voids [203] & 0.15\\
\midrule
$\left[10.5; 12\right]$ &  Nodes [138] & Voids [130] & $\mathbf{10^{-3}}$ & Nodes [95]& Voids [114] & 0.19\\
$\left[10.5; 12\right]$ &  Nodes [138] & Filaments [245] & \textbf{0.02}&   Nodes [95]& Filaments [201] & 0.69 \\
$\left[10.5; 12\right]$ &  Filaments [245] & Voids [130]  & $\mathbf{10^{-2}}$ &  Filaments [201] & Voids [114] & \textbf{0.01}\\
\bottomrule
\end{tabular}
\label{PDFlambdaResults}
\end{table*}

\begin{figure*}
\includegraphics[width=1.0\textwidth]{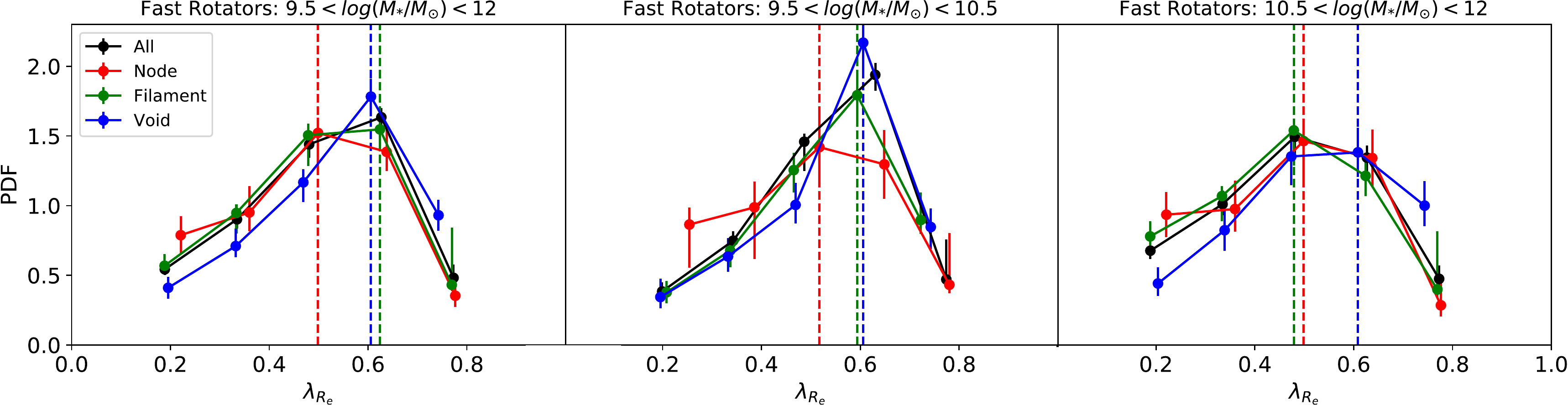}
\caption{PDFs of the $\lambda_{\rm R_e}$ distributions for 976 SAMI fast rotators (black) in nodes (red), filaments (green) and voids (blue). The left, middle, and right panels refer to galaxies with $9.5<\log(M_{\star}/M_{\odot})<12$, $9.5<\log(M_{\star}/M_{\odot})<10.5$, and $10.5<\log(M_{\star}/M_{\odot})<12$, respectively. The dashed lines mark the peaks for the PDFs of the $\lambda_{\rm R_e}$ distributions.}
\label{PDFlambdaFR}
\end{figure*}

To avoid the bias introduced by physical processes that remove angular momentum and produce SRs, we select 976 FRs out of the initial sample of 1094 SAMI galaxies. This allows us to look for possible segregations between the spin properties of rotation-dominated galaxies in different large-scale environments. Figure~\ref{PDFlambdaFR} shows the PDFs for 976 FRs in nodes, filaments, and voids, for the whole mass range (left panel), low-mass galaxies (middle panel) and high-mass galaxies (right panel). Comparing Figure~\ref{PDFlambdaFR} with Figure~\ref{PDFlambda}, the removal of SRs decreases the number of galaxies with low $\lambda_{\rm R_e}$ in nodes. The $\lambda_{\rm R_e}$ distributions show an approximately lognormal shape for the PDF of spin amplitudes, as expected from the tidal torque theory and seen for dark matter halos in simulations \citep{GaneshaiahVeena2018}.

The number of FRs for each subsample and the $p$-values from the two-sample K-S test are reported in columns~5--7 of Table~\ref{PDFlambdaResults}. For the whole sample of FRs, significant differences are found between the $\lambda_{\rm R_e}$ distributions for galaxies in voids relative to those in both nodes and filaments. For low-mass FRs, only the $\lambda_{\rm R_e}$ PDFs for galaxies in voids and nodes are kinematically distinct, while for high-mass FRs, only the $\lambda_{\rm R_e}$ PDFs for voids and filaments are significantly different. 

Table~\ref{PDFlambdaResultsFR} lists the mode (peak of the PDF) and median values of $\lambda_{\rm R_e}$ for FRs belonging to each type of large-scale structure. The modes are in all cases higher for void galaxies ($\lambda_{\rm R_e}$\,$\approx$\,0.6) than for node galaxies ($\lambda_{\rm R_e}$\,$\approx$\,0.5), with filament galaxies more similar to void galaxies for the whole sample and low-mass galaxies, and more similar to node galaxies for the high-mass sample. This is to say that galaxies in nodes tend to be more dispersion-dominated than those in voids, while those in filaments resemble void galaxies at low masses and node galaxies at high masses.

Overall, galaxies in different cosmic web structures show evidence for kinematic segregation; this evidence is stronger when SRs are included, but is still present for FRs only. Galaxies in voids and low-mass galaxies in filaments tend to have higher $\lambda_{\rm R_e}$ than galaxies in nodes and high-mass galaxies in filaments. These results demonstrate that the amplitude of a galaxy's spin, and not only its orientation, is related to its large-scale environment.

\begin{table}
\caption{Mode (peak) and median values for the $\lambda_{\rm R_e}$ of fast rotators in nodes, filaments, and voids. Columns~1 gives the $M_{\star}$ range, while columns~2,~3 and~4 report the $\lambda_{\rm R_e}$ mode and median for nodes, filaments, and voids, respectively.}
\centering
\begin{tabular}{crccc}
\toprule
$M_{\star}$ range &  & Nodes  & Filaments &  Voids\\
\midrule
$\left[9.5; 12.0\right]$ & $\lambda_{\rm R_{e}, peak}$ : & 0.50  & 0.62 & 0.61\\
& $\langle\lambda_{\rm R_{e}}\rangle$ : & 0.470$\pm$0.007 & 0.501$\pm$0.004 &  0.568$\pm$0.009\\
\midrule
$\left[9.5; 10.5\right]$& $\lambda_{\rm R_{e}, peak}$ :& 0.52  &  0.59 & 0.61   \\
& $\langle\lambda_{\rm R_{e}}\rangle$ : &  0.501$\pm$0.009 & 0.544$\pm$0.007 & 0.576$\pm$0.010\\
\midrule
$\left[10.5; 12.0\right]$  &$\lambda_{\rm R_{e}, peak}$ : & 0.50 & 0.48 &0.61   \\
& $\langle\lambda_{\rm R_{e}}\rangle$ : & 0.462$\pm$0.010 & 0.475$\pm$0.009 & 0.537$\pm$0.020\\
\bottomrule
\end{tabular}
\label{PDFlambdaResultsFR}
\end{table}

\subsection{Correlations of \texorpdfstring{$\lambda_{\rm R_e}$}{} with galaxy internal properties and local and large-scale environment metrics}
\label{Correlations}

To assess the impact of the large-scale environment on stellar spin in more detail, we investigate the dependency of $\lambda_{\rm R_e}$ on the distances to the closest node, filament, and void. We use the Spearman rank correlation test (correlation coefficient, $\rho$, and $p$-value, $p_{\rm S}$) to quantify the correlation between two variables and its statistical significance. For comparison with previous studies that focus on assessing the correlations between $\lambda_{\rm R_e}$ and internal galaxy properties or local environment, we also quantify the dependencies of $\lambda_{\rm R_e}$ on stellar mass, age, and local galaxy density for the same SAMI sample. 

Since in the past the strongest trends have been found
for high-mass galaxies relative to low-mass galaxies (e.g., \citealp{vandeSande2021b,Croom2024}), we report correlations for the whole SAMI sample of 1062 galaxies (see Section~\ref{Galaxy sample}), for 294 low-mass galaxies with $9.5<\log(M_{\star}/M_{\odot})<10.5$ and for 768 high-mass galaxies with $10.5<\log(M_{\star}/M_{\odot})<12$. 

\begin{figure*}
\includegraphics[width=1.0\textwidth]{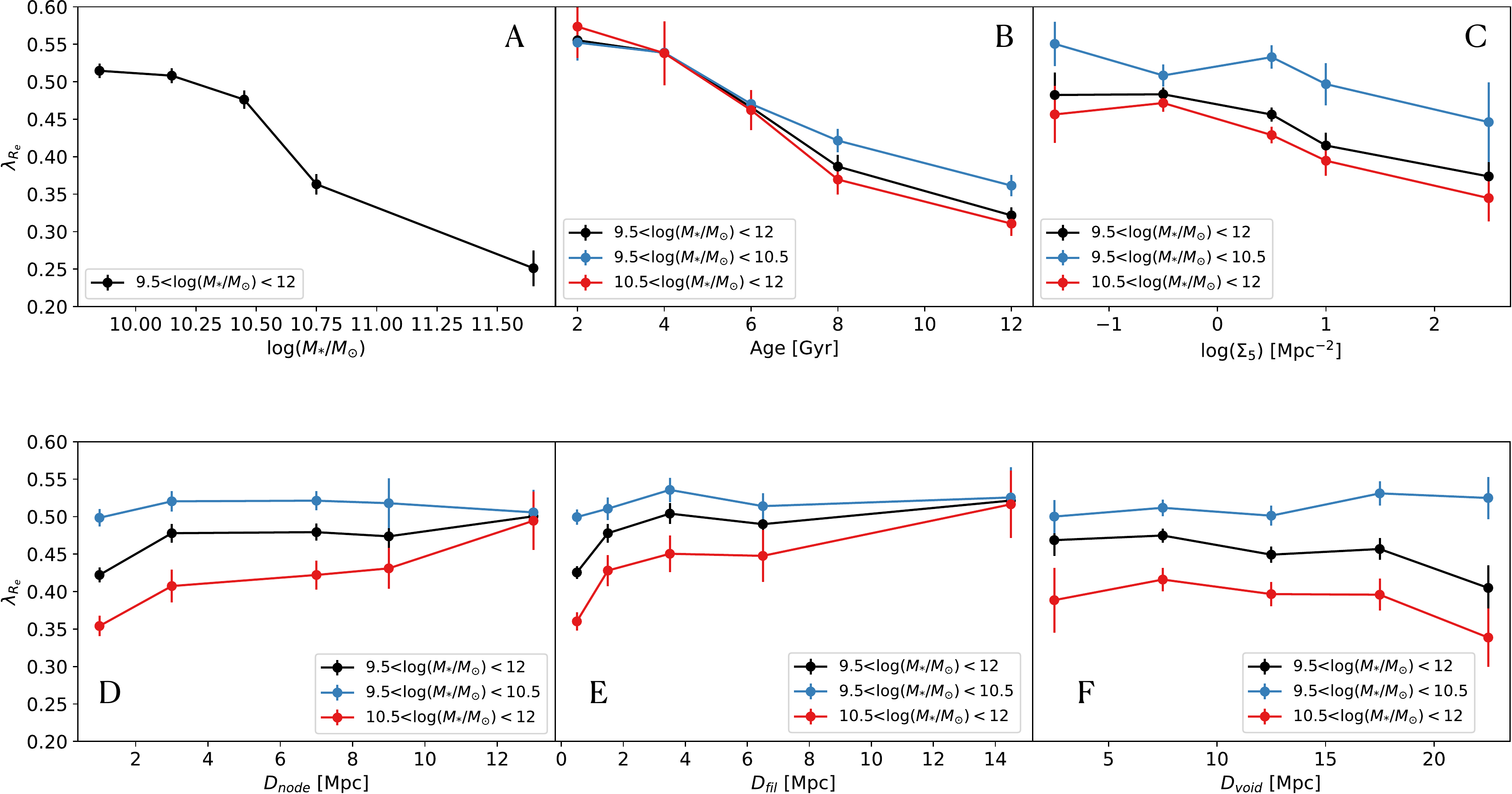}
\caption{Average $\lambda_{\rm R_e}$ values in bins of stellar mass (panel~A), stellar age (panel~B), local galaxy density (panel~C), distance to node (panels D), distance to filament (panel~E), and distance to void (panel~F). Standard errors on the mean are shown.}
\label{SpinvsGalaxyParameters}
\end{figure*}

Figure~\ref{SpinvsGalaxyParameters} shows average $\lambda_{\rm R_e}$ in bins of stellar mass (panel~A), stellar age (panel~B), local galaxy density (panel~C), distance to node (panel~D), distance to filament (panel~E), and distance to void (panel~F). The mean $\lambda_{\rm R_e}$ decreases with increasing $M_\star$, age, $\Sigma_{5}$ and $D_{\rm void}$, and increases for higher values of $D_{\rm node}$ and $D_{\rm fil}$. Table~\ref{SpearmanResults} reports the results from the Spearman test. The strongest correlations are found for the internal galaxy properties, with stellar age being the dominant parameter. Of the environment metrics, $\lambda_{\rm R_e}$ has the strongest dependence on $D_{\rm fil}$ ($\rho=0.18$, $p_{\rm S}=10^{-9}$). Both $\Sigma_{5}$ ($\rho=-0.16$, $p_{\rm S}=10^{-7}$) and $D_{\rm node}$ ($\rho=0.14$, $p_{\rm S}=10^{-6}$) also show strong correlations, while $D_{\rm void}$ ($\rho=-0.07$, $p_{\rm S}=0.018$) has the weakest dependency. Low-mass galaxies have higher $\lambda_{\rm R_e}$ than high-mass galaxies, with both populations showing similar trends as a function of the analysed properties, although stronger for the high-mass sample.

\begin{figure*}
\centering
\includegraphics[scale=0.28]{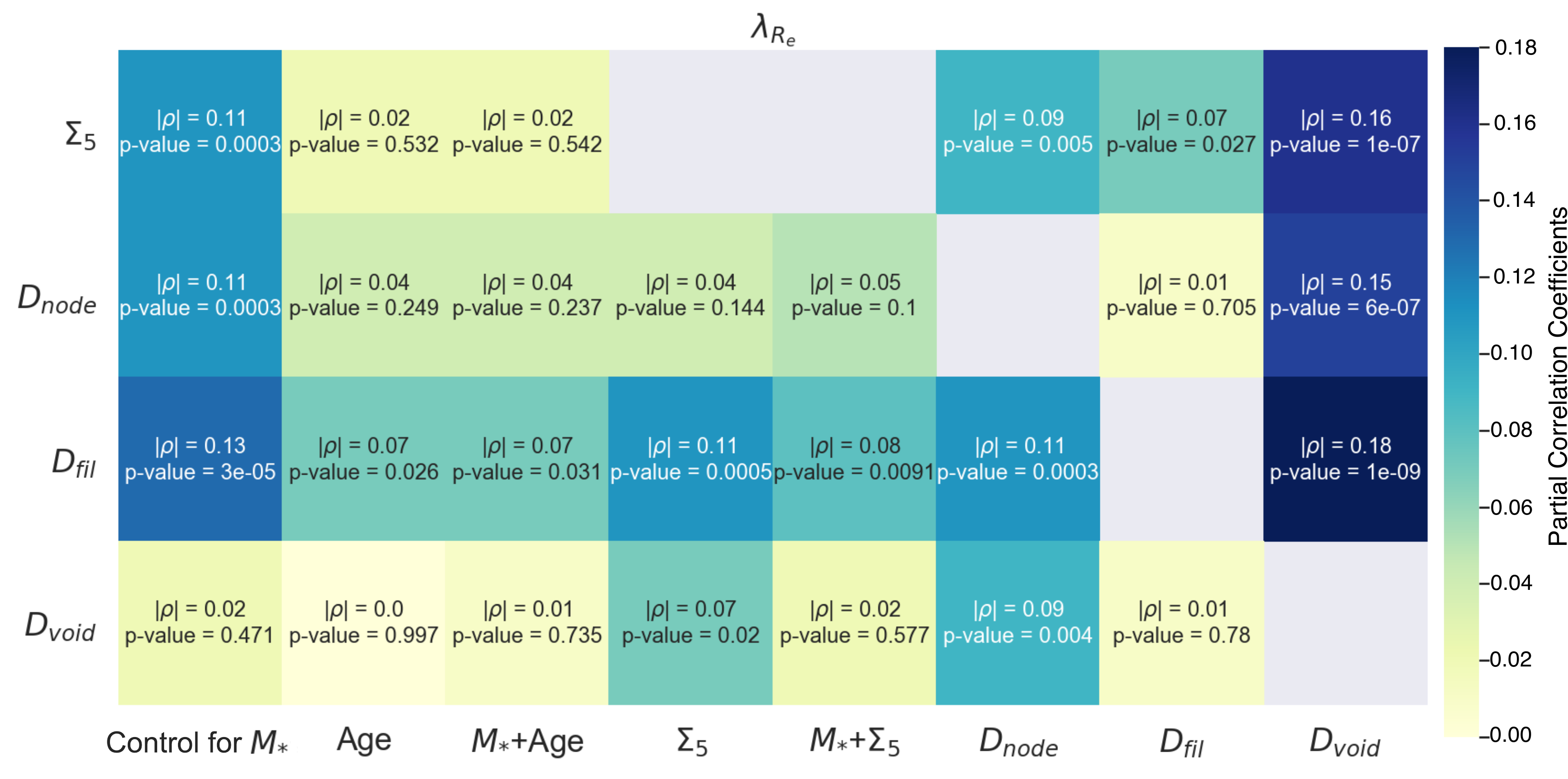}
\caption{Partial correlation coefficients from the Spearman test between $\lambda_{\rm R_e}$ and environment metrics while controlling for $M_{\star}$, age, $M_{\star}$+age, $\Sigma_5$, $M_{\star}$+$\Sigma_5$, $D_{\rm node}$, $D_{\rm fil}$ and $D_{\rm void}$.}
\label{PartialCorrelations}
\end{figure*}

To explore the true correlation between two parameters while controlling a third quantity (avoiding the cross-correlation driven by their dependency on the third property), we estimate the partial correlation coefficients for $\lambda_{\rm R_e}$ \citep{Lawrance1976}. We control for $M_\star$, age, $\Sigma_{5}$, $D_{\rm node}$, $D_{\rm fil}$ and $D_{\rm void}$. We also control for $M_\star$ and age at the same time (since they are the strongest internal properties driving $\lambda_{\rm R_e}$), and for $M_{\star}$ and $\Sigma_{5}$ simultaneously (since they are strong indirect tracers of the cosmic web). The results are reported in Figure~\ref{PartialCorrelations}. Controlling for $M_\star$ shows strong residual correlations for all environment metrics, while controlling for age leaves only a marginally significant result for $D_{\rm fil}$ ($p$-value $=0.026$). The results for $M_\star$+age are consistent with those controlling only for age, leaving a marginally significant result for $D_{\rm fil}$ ($p$-value $=0.031$). This highlights that age is the dominant correlation parameter with $\lambda_{\rm R_e}$ \citep{Croom2024}, but $D_{\rm fil}$ plays an independent role in shaping stellar spin. No correlation with $D_{\rm node}$ remains while controlling for $\Sigma_{5}$ or $D_{\rm fil}$. A strong correlation is left for $\Sigma_{5}$ while controlling for $D_{\rm node}$ ($p$-value $=0.005$), but it is more marginal while controlling for $D_{\rm fil}$ ($p$-value $=0.027$). Controlling for $M_\star$+$\Sigma_{5}$ leaves significant correlation of residuals only for $D_{\rm fil}$ ($p$-value $=0.009$). A complementary analysis of kinematic residuals in $D_{\rm fil}$ after accounting for $M_\star$ and $\Sigma_{5}$ is shown in Appendix~\ref{Kinematic residuals}, reaching the same conclusion.
The strong residuals on all metrics while controlling for $D_{\rm void}$ confirm the weak
dependency of $\lambda_{\rm R_e}$ on this parameter.

Overall, $D_{\rm fil}$ is found to be the most significant environment metric correlating with $\lambda_{\rm R_e}$, although it is secondary to the dominant role played by the stellar population age of the galaxy. This result points to the stage in the formation of cosmic web structures that has the most impact on galaxy spin.

\begin{table}
\centering
\caption{Spearman rank correlation tests for stellar spin $\lambda_{\rm R_e}$ as a function of internal galaxy properties, local environment metrics, and cosmic web metrics. Column~1 gives the galaxy property analysed, column~2 the selected stellar mass range, column~3 the number of galaxies in the sample, column~4 the Spearman correlation coefficient, and column~5 the Spearman test $p$-value (in bold if less than 0.05).}
\label{SpearmanResults}
\begin{tabular}{@{}lcccc@{}}
\toprule
Galaxy property & $M_{\star}$ range & $N_{\rm gal}$ & $\rho$  & $p_{\rm S}$  \\
\toprule
$\log(M_\star/M_{\odot})$ &  [9.5; 12]  & 1062 & $-$0.31 & $\mathbf{10^{-25}}$\\
Age        &  [9.5; 12]  & 1062 &    $-$0.48 &      $\mathbf{10^{-62}}$     \\
\midrule
$\Sigma_5$ &  [9.5; 12]  & 1062 & $-$0.16 & $\mathbf{10^{-7}}$\\
 & [9.5; 10.2] & 294 & $+$0.02 & 0.722\\
  & [10.2; 12] & 768 & $+$0.20 & $\mathbf{10^{-7}}$ \\
\midrule
$D_{\rm node}$ &  [9.5; 12]  & 1062 & $+$0.14 & $\mathbf{10^{-6}}$\\
 & [9.5; 10.2] & 294 & $+$0.10 & 0.098\\
  & [10.2; 12] & 768 & $+$0.15 & $\mathbf{10^{-5}}$\\
$D_{\rm fil}$&  [9.5; 12]  & 1062 &$+$0.18 & $\mathbf{10^{-9}}$\\
 & [9.5; 10.2] & 294 & $+$0.08 & 0.166\\
  & [10.2; 12] & 768 & $+$0.20 & $\mathbf{10^{-8}}$\\
$D_{\rm void}$&  [9.5; 12]  & 1062 & $-$0.07 & \textbf{0.018}\\
 & [9.5; 10.2] & 294 & $+$0.07 &0.203 \\
  & [10.2; 12] & 768 & $+$0.07 & \textbf{0.045}\\
\bottomrule
\end{tabular}
\end{table}

\begin{table}
\centering
\caption{Median values for SR, FR and $M_{\star}$--matched FR distributions in stellar mass, age, local galaxy density, halo mass, local type, and distances to the nearest node, filament and void.}
\label{MediansResults}
\begin{tabular}{@{}lccc@{}}
\toprule
Galaxy property & FR & SR & $M_{\star}$--matched FR \\
\toprule
$\log(M_\star/M_{\odot})$ & 10.42$\pm$0.01 & 11.03$\pm$0.04 & 11.03$\pm$0.04  \\
Age [Gyr]       &    5.42$\pm$0.12 &  11.11$\pm$0.24 & 11.14$\pm$0.25  \\
$\log(\Sigma_5$) [Mpc$^{-2}$] & 0.13$\pm$0.02 & 0.78$\pm$0.07 & 0.69$\pm$0.07 \\
$\log(M_{\rm halo}/M_{\odot})$ & 13.27$\pm$0.03 & 13.50$\pm$0.06 & 13.53$\pm$0.06\\
local type &  satellite & central & central \\
$D_{\rm node}$ [Mpc]& 3.06$\pm$0.11 & 0.89$\pm$0.27 & 1.53$\pm$0.35\\
 $D_{\rm fil}$ [Mpc]& 0.99$\pm$0.07 & 0.28$\pm$0.18 & 0.42$\pm$0.21 \\
 $D_{\rm void}$ [Mpc]&  10.69$\pm$0.17 & 13.60$\pm$0.60 & 11.14$\pm$0.56 \\
 \bottomrule
\end{tabular}
\end{table}

\subsection{Histograms of FRs and SRs}
\label{Histograms of SRs and FRs}

We next explore the spin distributions of 950 FRs and 105 SRs as a function of internal galaxy properties, local environment metrics and large-scale environment metrics. The median values of each property for the SR and FR distributions are listed in Table~\ref{MediansResults}. SRs are generally more massive and older than FRs, and they tend to populate higher-density regions, to be in massive groups, and to be central galaxies. SRs tend to be closer to nodes and to filaments, while FRs tend to be closer to void regions. According to the two-sample K-S test, the most significant differences between the FR and SR distributions are found for their stellar mass and age ($p_{\rm 2 K-S}=10^{-15}$). Among the environment metrics, the local type and distance to filament show the strongest differences ($p_{\rm 2 K-S}=10^{-8}$). Similar results are found for high-mass FRs and high-mass SRs with $\log(M_{\star}/M_{\odot})>10.2$, where the identification of SRs is less ambiguous (e.g., \citealp{vandeSande2021a}).

\begin{figure*}
\includegraphics[scale=0.33]{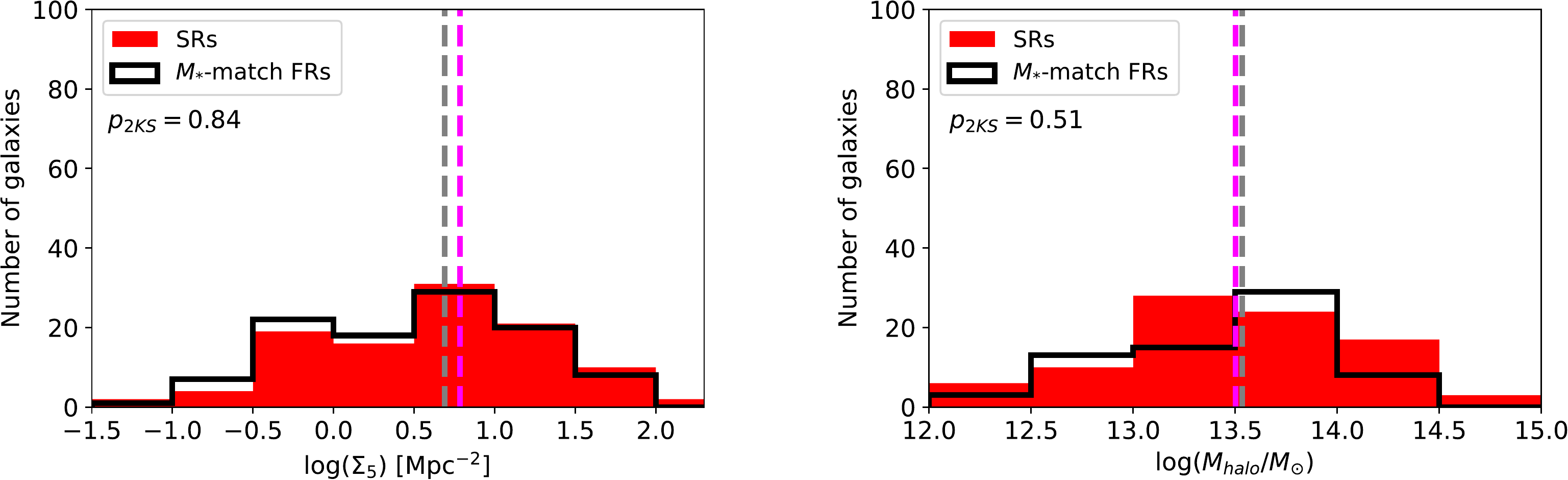}
\vspace{2.5mm} \\
\includegraphics[scale=0.33]{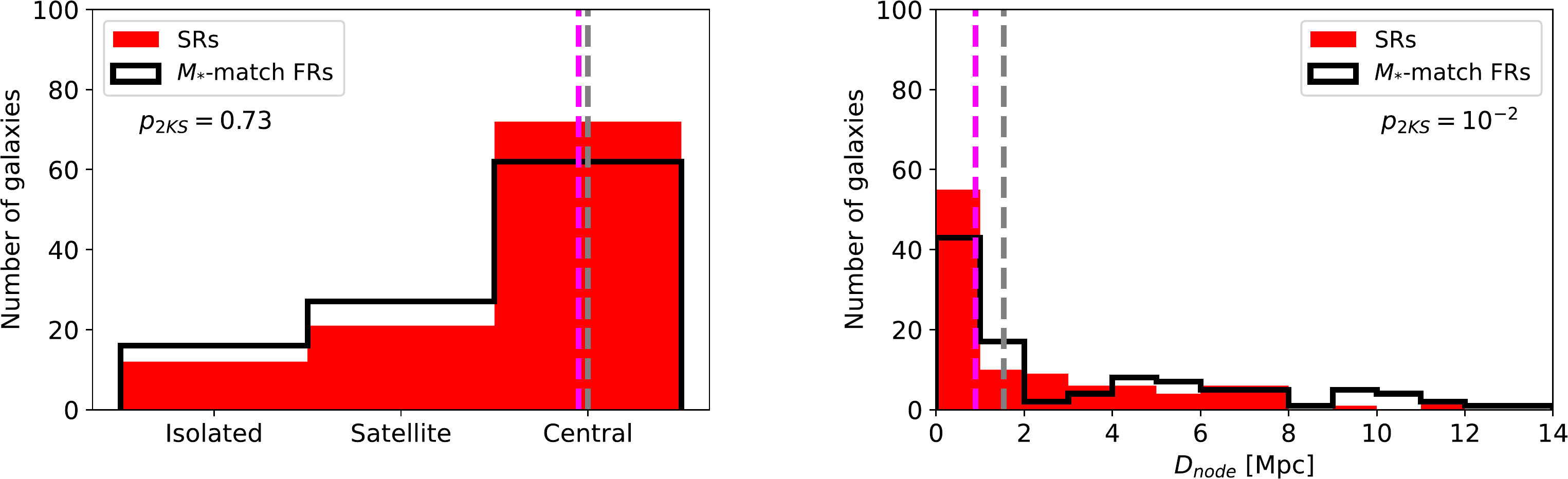}
\vspace{2.5mm} \\
\includegraphics[scale=0.33]{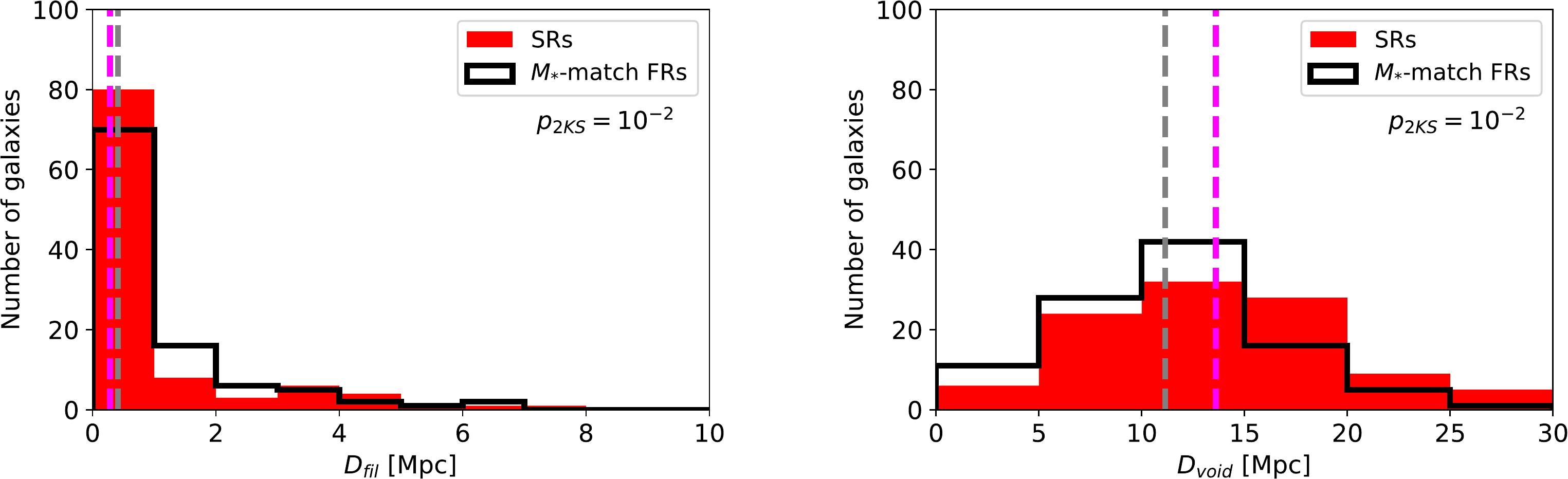}
\caption{Distributions of local galaxy density, halo mass, local type, and distances to the nearest node, filament and void for SRs (red) and $M_{\star}$-matched FRs (black). The dashed lines mark the median values for SRs (magenta) and $M_{\star}$-matched FRs (grey). Two-sample K-S test $p$-values are reported in each panel.}
\label{histogramsFR_SR_match}
\end{figure*}

To exclude the effects of the internal galaxy parameters, which can influence the correlations with environmental metrics, we explore the differences between SRs and a sample of stellar-mass-matched FRs. To build the latter sample we follow a technique similar to that of \citet{Zovaro2024}, identifying the unique FR with the closest stellar mass to each SR. The matched samples contain 105 SRs and 105 $M_{\star}$-matched FRs. Figure~\ref{histogramsFR_SR_match} shows the histograms for SRs and $M_{\star}$-matched FRs in local galaxy density, halo mass, local type, and distances to the nearest node, filament, and void. According to the two-sample K-S test, there are significant differences between the distributions of the large-scale environment metrics for the SR and $M_{\star}$-matched FR samples. On the other hand, the two samples are consistent with belonging to the same population in terms of local environment, halo mass, and central/satellite/isolated classification. 

The same conclusions are found if we compare SRs with a sample of age-matched FRs. Overall, these results are in agreement with the conclusions from the partial correlation analysis of previous Section~\ref{Correlations}. Both analyses show that the cosmic web, especially filaments, has a role in determining galaxy spins, apparent even after accounting for the dominant effects of stellar mass, age, local galaxy density and simultaneous combinations of these galaxy properties.

\subsection{Distribution of SRs in the large-scale environment}
\label{Fraction of SRs in environmental planes}

To better understand the role of local and large-scale environment in the formation of SRs, we investigate the distribution of FRs and SRs in the planes defined by pairs of environmental parameters.

\subsubsection{Large-scale environment plane}

\begin{figure}
\includegraphics[width=1.1\columnwidth]{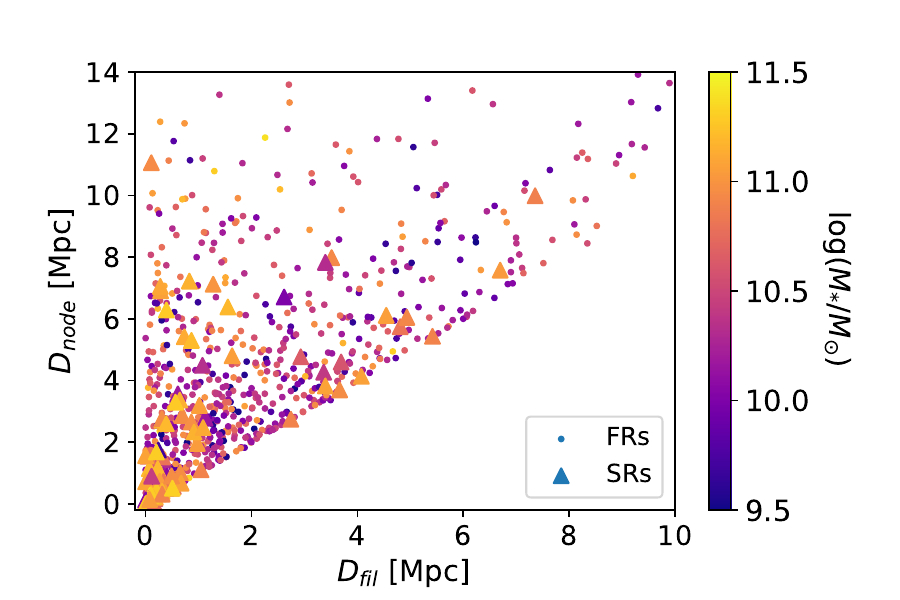}
\includegraphics[width=1.1\columnwidth]{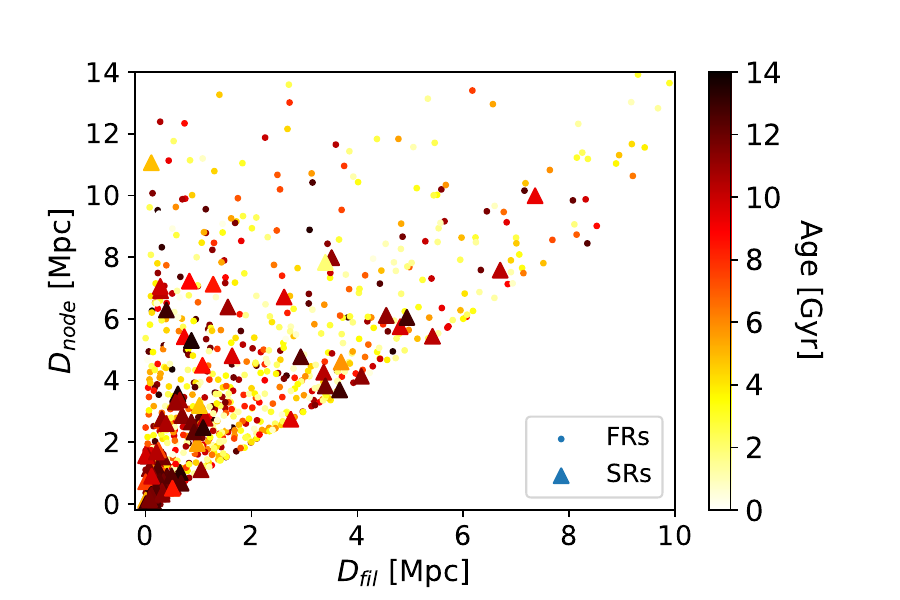}
\includegraphics[width=1.1\columnwidth]{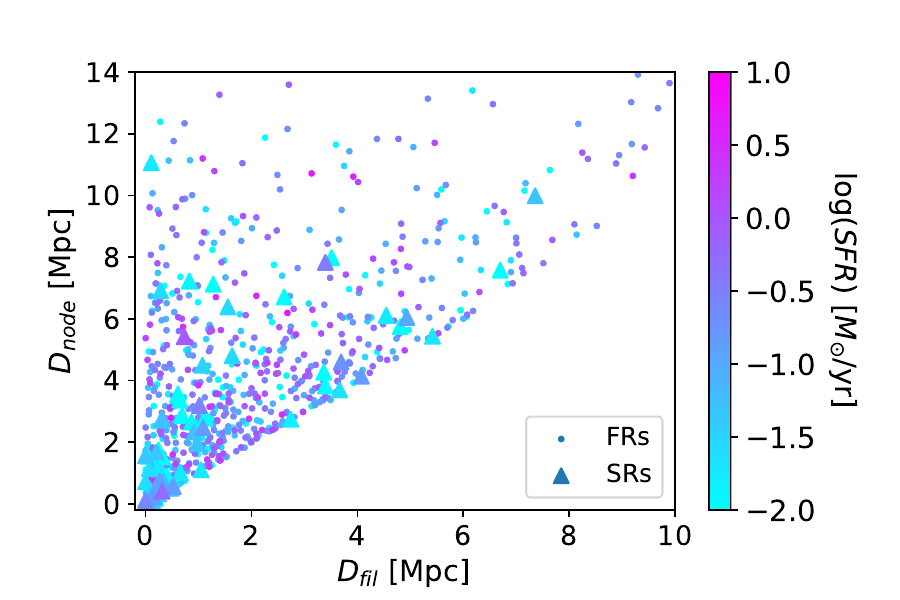}
\caption{$D_{\rm fil}$--$D_{\rm node}$ plane for 950 FRs (circles) and 105 SRs (triangles) coloured by stellar mass (top), age (middle), and star formation rate (bottom).}
\label{DfilDnode_plane}
\end{figure}

Figure~\ref{DfilDnode_plane} shows the $D_{\rm fil}$--$D_{\rm node}$ plane for 950 FRs (circles) and 105 SRs (triangles) colour-coded by stellar mass (top), age (middle) and star formation rate (bottom). Half the plane (below/right of the bisector) is empty because $D_{\rm fil} \leq D_{\rm node}$ by construction (Section~\ref{Large-scale environment metrics}). FRs are scattered across the remaining plane, while SRs tend to be found closer to the nearest filament and node. SRs far from filaments or nodes are still massive and old, with low SFR. Such SRs might be members of groups who have not been identified because of their low membership, or because they are in groups or filaments that have not been sampled by the GAMA survey.

\begin{figure}
\includegraphics[width=1.1\columnwidth]{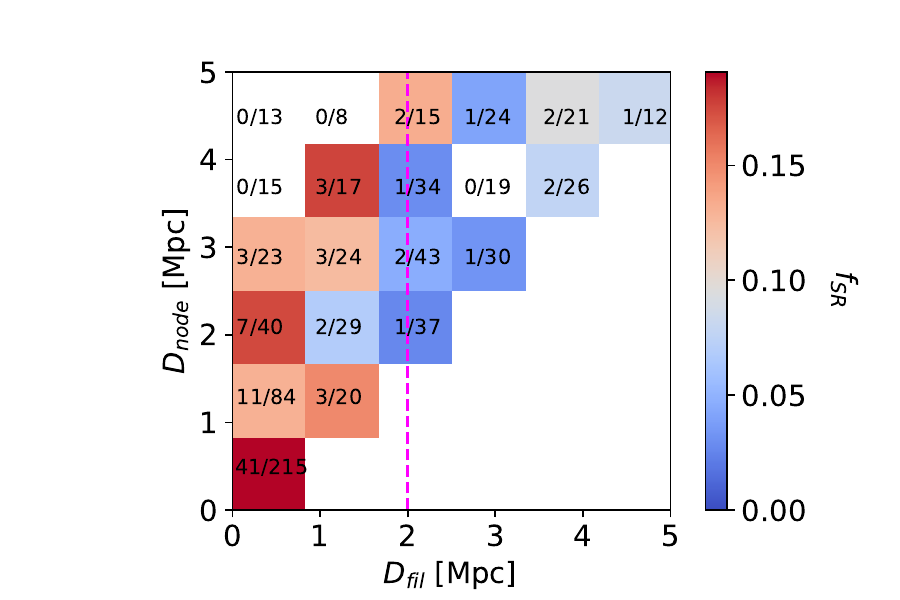}
\caption{Fractions of SRs in bins over the zoomed-in $D_{\rm fil}$--$D_{\rm node}$ plane within $5$\,Mpc. The regions closer to filaments and nodes have higher $f_{\rm SR}$ (redder colours), while regions at large distances from filaments and nodes have lower $f_{\rm SR}$ (bluer colours). About 95\% of SRs can be found within $D_{\rm fil}\leq2$\,Mpc (magenta dashed line).}
\label{Dfil_Dnode_fSR}
\end{figure}

Zooming in on the $D_{\rm fil}$--$D_{\rm node}$ plane for galaxies within $5$\,Mpc of filaments and nodes, Figure~\ref{Dfil_Dnode_fSR} shows the SR fraction, $f_{\rm SR}=N_{\rm SR}/N_{\rm (SR+FR)}$, in bins. The regions closer to filaments and nodes are characterised by higher values of $f_{\rm SR}$ (redder colours), whereas $f_{\rm SR}$ is lower (bluer colours) for the regions at large distances from filaments and nodes. Almost all SRs ($\sim$95\%) can be found within $D_{\rm fil}\leq2$\,Mpc, while only $\sim$60\% of FRs have $D_{\rm fil}\leq2$\,Mpc. Both SRs and FRs span a wide range of distances from nodes.

\begin{figure*}
\includegraphics[scale=0.30]{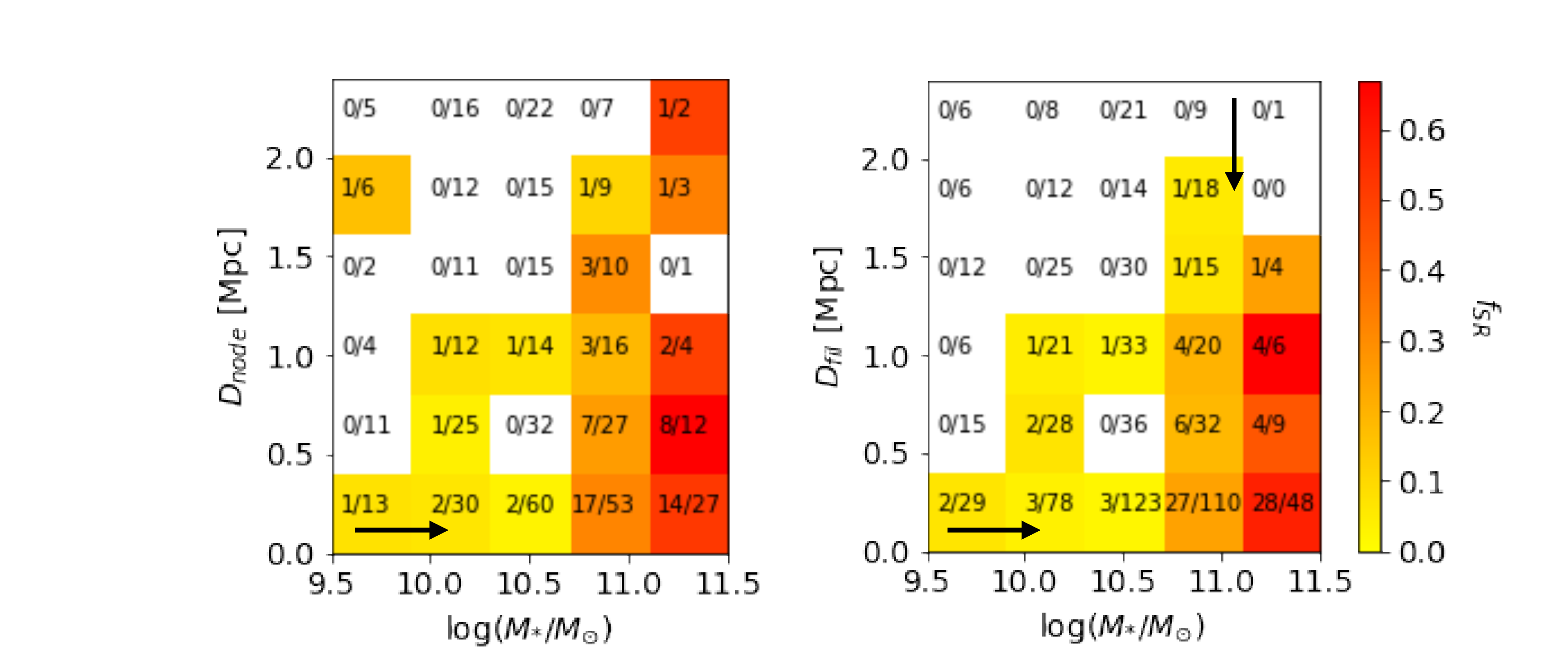}
\caption{The fraction of galaxies that are SRs, $f_{\rm SR}$, as a joint function of stellar mass and distance to the closest node (left) or filament (right). For each bin, we indicate the number of SRs to
the total SRs+FRs. Redder colours correspond to higher $f_{\rm SR}$. The black arrows show the $f_{\rm SR}$ trends with each property.}
\label{Dfil_Dnode_Mass_fSR}
\end{figure*}

To understand the impact of the large-scale environment on SRs while taking stellar mass into account, we reproduce Figure~8 of \citet{vandeSande2021a}, who investigated mass and local galaxy density. Figure~\ref{Dfil_Dnode_Mass_fSR} displays the fraction of SRs as a joint function of mass and distance to the closest node (left panel) or filament (right panel), for distances $<$2.5\,Mpc. At fixed $D_{\rm node}$ or $D_{\rm fil}$, $f_{\rm SR}$ increases for higher stellar masses. At fixed $M_{\star}$, $f_{\rm SR}$ increases for smaller $D_{\rm fil}$ for galaxies with $\log(M_{\star}/M_{\odot})>10.7$. The trend is stronger for $D_{\rm fil}$, where the great majority of SRs are found at $D_{\rm fil}<0.5$\,Mpc, while it is not significant for $D_{\rm node}$. This highlights the role played by filaments in the formation of SRs.

\subsubsection{Local and large-scale environmental planes}

We study the distribution of 950 FRs and 105 SRs in environmental planes pairing local and large-scale environment metrics (i.e.\ joint distributions over one local environment metric and one large-scale environment metric). This allows us to shed light on the type of environment in which SRs tend to be formed.

\begin{figure*}
\includegraphics[scale=0.38]{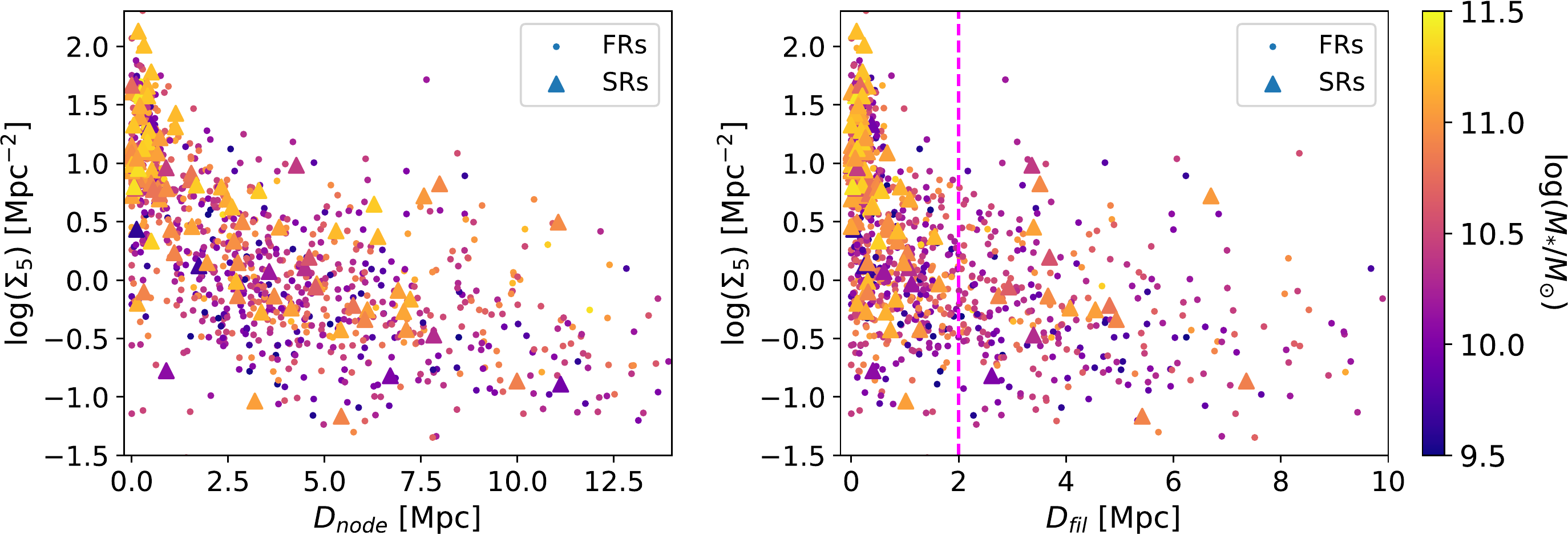}
\vspace{0.3cm}
\includegraphics[scale=0.38]{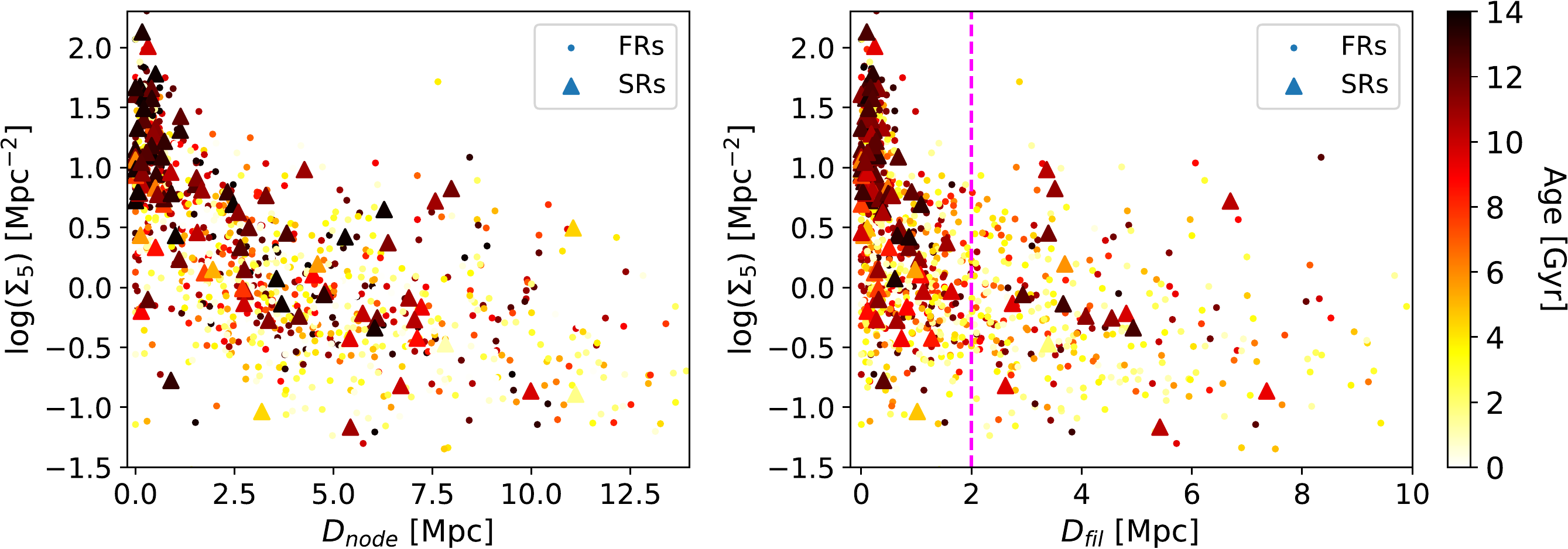}
\vspace{0.3cm}
\includegraphics[scale=0.38]{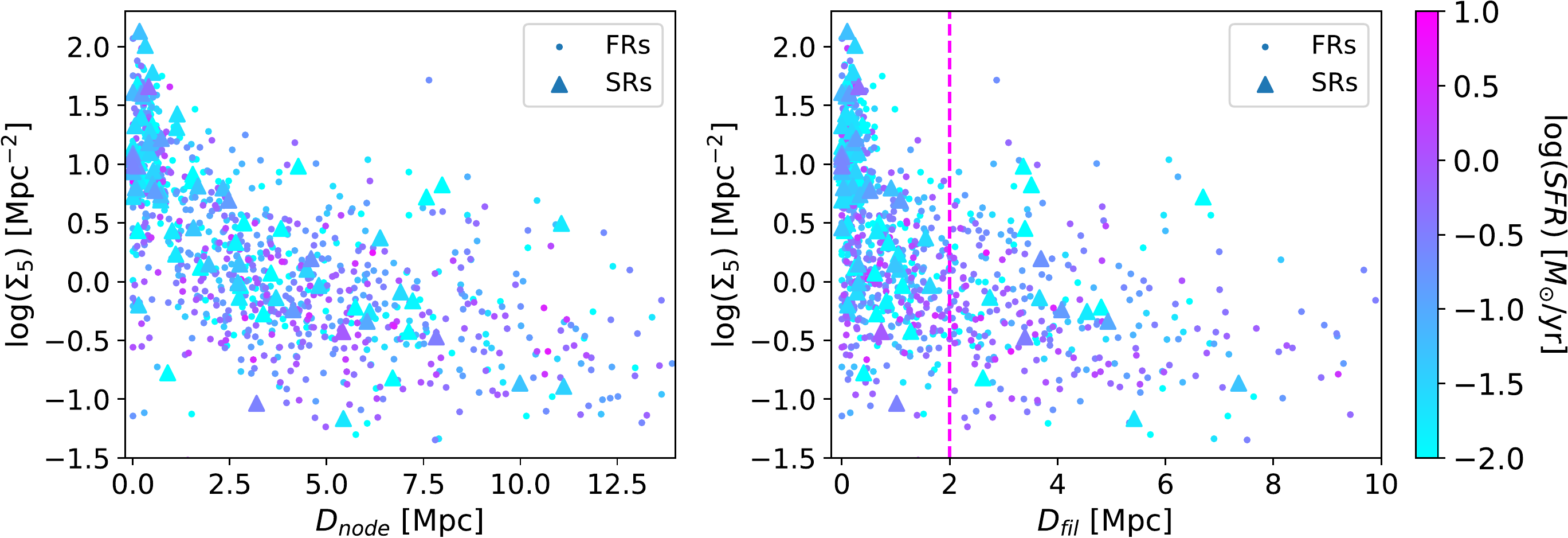}
\caption{Distribution in the $\Sigma_{\rm 5}$--$D_{\rm node}$ plane (left) and $\Sigma_{\rm 5}$--$D_{\rm fil}$ plane (right) of 950 FRs (circles) and 105 SRs (triangles), colour-coded by stellar mass (top), age (middle) and star formation rate (bottom). About 95\% of SRs are concentrated at $D_{\rm fil}<2$\,Mpc (dashed magenta line), but they cover almost the full range of local galaxy densities spanned by FRs.}
\label{Sigma_DfilDnode_plane}
\end{figure*}

Figure~\ref{Sigma_DfilDnode_plane} shows the joint distribution over local galaxy density $\Sigma_{\rm 5}$ and distance to the nearest node, $D_{\rm node}$ (left), or filament, $D_{\rm fil}$ (right), with galaxies colour-coded by stellar mass (top), age (middle), and SFR (bottom). SRs can be found scattered over the whole $\Sigma_{\rm 5}$--$D_{\rm node}$ plane in nearly the same way as FRs, although the most massive and oldest SRs tend to be found in high-density environments and are closer to nodes: $\sim$50\% of SRs have $D_{\rm node}<1.5$\,Mpc and $\log\Sigma_{\rm 5}>0.5$\,Mpc$^{-2}$. The SRs at larger distances from nodes and filaments, and in low-density environments, tend to have low SFR. In the $\Sigma_{\rm 5}$--$D_{\rm fil}$ plane, about 95\% of SRs are concentrated at small distances from filaments, $D_{\rm fil}<2$\,Mpc, but they cover almost the full range of local galaxy densities spanned by FRs. This is confirmed by the two-sample K-S test since the $\Sigma_{\rm 5}$ distributions within $D_{\rm fil}<2$\,Mpc for SRs and FRs are not significantly different.

\begin{figure*}
\includegraphics[scale=0.35]{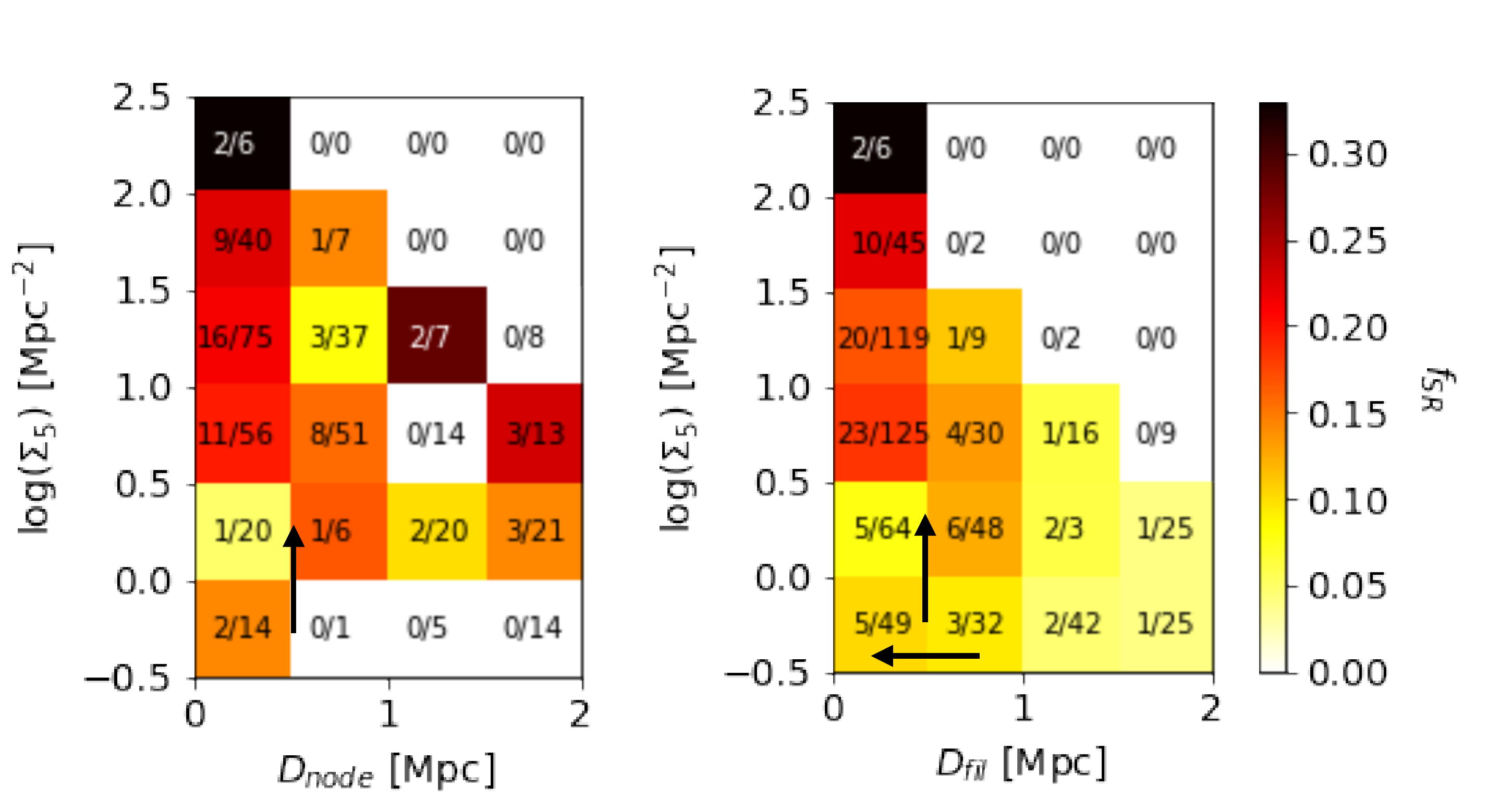}
\caption{Fractions of SRs as a joint function of local galaxy density and distance to the closest node (left) or filament (right). For each bin, we indicate the numbers of
SRs and SRs+FRs. The black arrows show that at both fixed $D_{\rm node}<0.5$\,Mpc and $D_{\rm fil}<0.5$\,Mpc, the fraction of SRs increases for higher local galaxy density. At fixed
$\Sigma_5$, $f_{\rm SR}$ increases for shorter $D_{\rm fil}$ values for all galaxies.}
\label{Sigma_Dnode_fil_fSR}
\end{figure*}

To investigate whether there is a trend with local environment at fixed large-scale environment, Figure~\ref{Sigma_Dnode_fil_fSR} shows the fraction of SRs as a joint function of local galaxy density and distance (within 2\,Mpc) to either the closest node (left) or the closest filament (right). At fixed $D_{\rm node}<0.5$\,Mpc or fixed $D_{\rm fil}<0.5$\,Mpc, the fraction of SRs increases for higher local galaxy density. At larger distances from nodes and filaments, there is no clear trend with local environment. At fixed $\Sigma_{5}$, $f_{\rm SR}$ increases for shorter $D_{\rm fil}$ values for all galaxies, whereas the trend is noisier for $D_{\rm node}$. This highlights again the importance for SR formation of $D_{\rm fil}$ relative to $D_{\rm node}$. Thus, we detect an influence of the local environment at fixed large-scale environment only for galaxies very close to filaments or nodes. Such influence may still be driven by stellar mass or age.

\begin{figure*}
\includegraphics[scale=0.38]{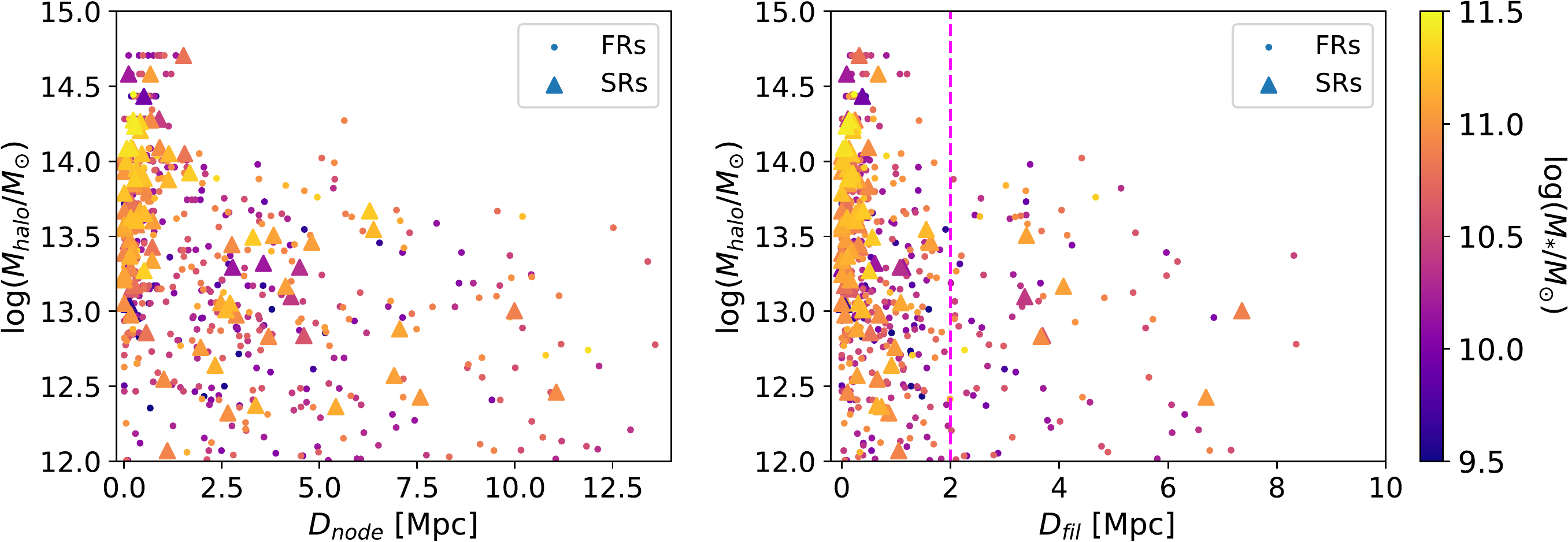}
\caption{$M_{\rm halo}$--$D_{\rm node}$ plane (left) and $M_{\rm halo}$--$D_{\rm fil}$ plane (right) for 545 FRs (circles) and 88 SRs (triangles), colour-coded by stellar mass. For $D_{\rm fil}<2$\,Mpc, SRs can be found from over nearly the same range of halo masses as FRs.}
\label{HaloMass_DfilDnode_plane}
\end{figure*}

Figure~\ref{HaloMass_DfilDnode_plane} shows the halo mass versus $D_{\rm node}$ (left) and $D_{\rm fil}$ (right) planes in stellar mass. There are 545 FRs and 88 SRs with halo mass measurements. Trends for SRs within this plane (and also in age and SFR) are similar to those in Figure~\ref{Sigma_DfilDnode_plane}: SRs are spread throughout the $M_{\rm halo}$--$D_{\rm node}$ plane, with the most massive, oldest, and lowest-SFR SRs in the most massive halos and closer to nodes. For $D_{\rm fil}<2$\,Mpc, SRs can be found from over nearly the same range of halo masses as FRs, with the distributions not significantly different according to the two-sample K-S test.

\begin{figure*}
\includegraphics[scale=0.33]{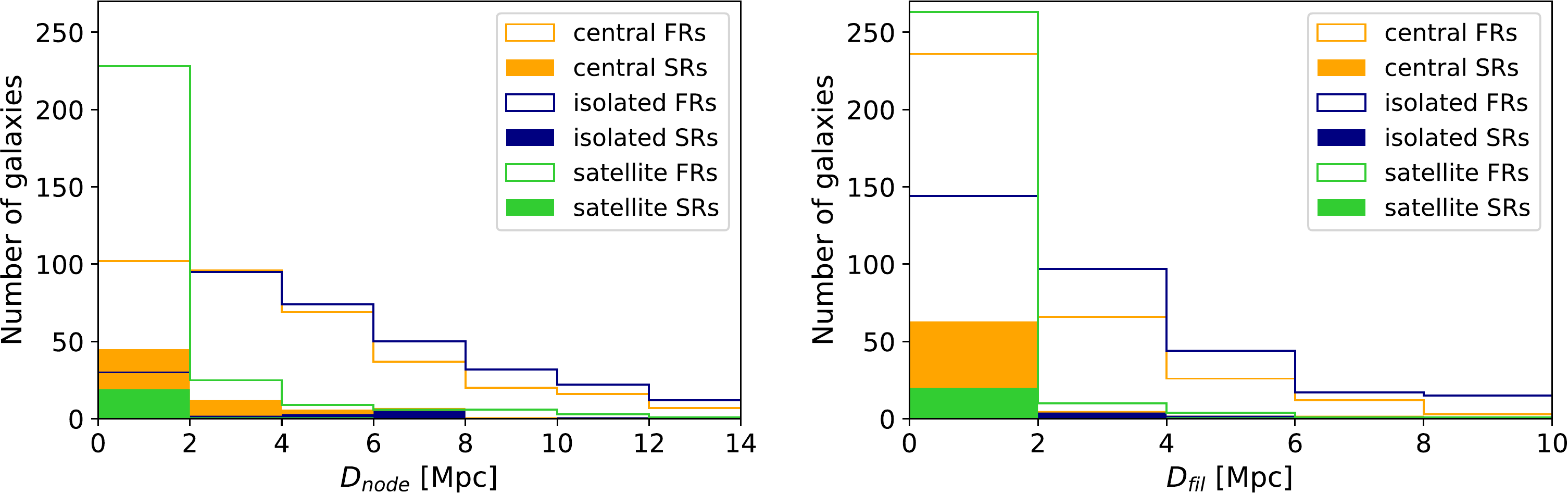}
\caption{Local type histograms of 105 SRs and 950 FRs for distances to closest node and filament. Most SRs are central galaxies having substantially smaller $D_{\rm fil}$ than FRs, while they cover large distances from nodes similar to FRs.}
\label{LocalType_DfilDnode_plane}
\end{figure*}

Finally, Figure~\ref{LocalType_DfilDnode_plane} shows the histograms of local type SRs and FRs for distances to closest node (left) and filament (right). Most SRs are central galaxies having substantially smaller $D_{\rm fil}$ than FRs, while (like FRs) they span a large range of distances from nodes.

Overall, the local versus large-scale environmental planes suggest that most SRs are located close to filaments, although they can be found over nearly as wide a range of local environments as FRs: low-density and high-density regions, low-mass and high-mass halos, and as both centrals and satellites. SRs that are close to nodes tend to be in extreme local environments: high-density regions and high-mass halos, where they are the central galaxies.

\section{Discussion}
\label{Discussion}

\subsection{Large-scale environment affects galaxy spin amplitudes}

In this work we investigate the connection between galaxy spin amplitudes and the large-scale structure of the cosmic web. Our results from Section~\ref{Spin amplitude} (Figure~\ref{PDFlambdaFR}) show that FRs in nodes tend to be more dispersion-dominated than those in filaments and voids, with the peaks of the $\lambda_{\rm R_e}$ distributions shifted for different cosmic web structures; these differences remain for high-mass and low-mass FRs. This outcome for galaxies is in agreement with the study of \citet{GaneshaiahVeena2018}, who used a cosmological N-body simulation to study the spin amplitudes of dark matter halos within large-scale structures. \citet{GaneshaiahVeena2018} use the spin parameter introduced by \citet{Bullock2001} as $\lambda=J/(\sqrt{2}MVR)$, where $J$ is the halo's angular momentum amplitude, $M$ is the halo mass, and $V$ is the halo circular velocity at radius $R$. They find that halos in filaments and walls have on average the highest spin (i.e.\ are more rotation-dominated), while halos in nodes show low $\lambda$ values (i.e. are more dispersion-dominated). Figure~8 of \citet{GaneshaiahVeena2018} shows that the halo spin distributions in all large-scale environments have lognormal shapes, but the peak values are shifted: from $\lambda=0.035$ in filaments and walls to $\lambda=0.030$ in voids and $\lambda=0.020$ in nodes. According to our findings, this relation between large-scale structure and spin amplitude for dark matter halos translates to galaxies, producing a relationship between stellar angular momentum and location in the cosmic web.

The analysis of the correlations of $\lambda_{\rm R_e}$ with galaxy internal properties, local environment metrics, and large-scale structures in Section~\ref{Correlations} reveals that, among the parameters explored in this study, stellar age is the primary correlate with spin amplitude. This agrees with the results of \citet{Croom2024}, who found that galaxy spin is more strongly correlated with age than with mass or local environment. \citet{Croom2024} point out that stellar age drives spin, but age itself is driven by the environment, so this latter does have an indirect effect on spin. In addition, we detect a significant residual correlation of $\lambda_{\rm R_e}$ with $D_{\rm fil}$, suggesting that the location of a galaxy with respect to the filamentary structure of the cosmic web also influences its stellar spin.

\subsection{Slow rotators are formed in filaments}

We investigate the distributions of fast rotators (FRs) and slow rotators (SRs) within the large-scale environment. While the locations of the two types overlap, SRs tend to be found closer to nodes and filaments than FRs, which tend to be found closer to voids. This difference also applies for SRs relative to stellar-mass-matched FRs, so it is not just a secondary correlation with mass: {\it large-scale} environment has a physical effect. On the other hand, no significant difference is detected in the distributions of SRs and FRs with {\it local} environment (Section~\ref{Histograms of SRs and FRs}). These findings point to a role for the cosmic web in the formation of SRs.  

From the analysis of the $D_{\rm fil}$--$D_{\rm node}$ plane in Section~\ref{Fraction of SRs in environmental planes}, $\sim$95\% of SRs have $D_{\rm fil}\leq2$\,Mpc. At fixed $M_{\star}$, the fraction of SRs significantly increases for decreasing $D_{\rm fil}$, especially for massive galaxies with $\log(M_{\star}/M_{\odot})>10.7$. This is in agreement with the finding that the spin of high-mass galaxies is more affected by environment than that of low-mass galaxies \citep{vandeSande2021b,Santucci2023,Croom2024}. The trends as a function of $D_{\rm node}$ are weaker. This suggests that pre-processing within filaments may be partially responsible for the formation of SRs before they reach the denser node environment. As a consequence, since the reduction in galaxy angular momentum begins at the filament scale, SRs are not required to be the central galaxies in groups, in agreement with the findings of \citet{Scuccimarra2024}, who investigated SRs in fossil groups versus non-fossil groups. 

The important role of filaments in the formation of SRs is in accord with the scenario from simulations where galaxy-galaxy mergers are the main cause of spin loss, after star formation quenching has depleted the gas reservoirs \citep{Lagos2018,Lagos2022}. Galaxy quenching occurs within filaments, where it is able to affect the gas contents \citep{Poudel2017,Kraljic2018,Hasan2023,Hoosain2024}. Mergers are more common physical processes within the less-dense filamentary structures, where galaxy relative velocities are lower, than in the more-dense nodes, where relative velocities are higher \citep{Dubois2014,Codis2015}. Thus, $D_{\rm fil}$ is a better tracer of mergers than $D_{\rm node}$, which correlates more strongly with extreme environmental processes such as dynamical friction, ram-pressure stripping, and strangulation (Barsanti et al. in preparation). The frequent detection of SRs in the central regions of galaxy clusters \citep{Cappellari2011,DEugenio2013,Houghton2013,Scott2014} might suggest that $D_{\rm node}$ should show more significant trends with $\lambda_{\rm R_e}$ and $f_{\rm SR}$ than $D_{\rm fil}$, but the difference in the physical processes traced by these two large-scale environment metrics explains why we see stronger correlations with $D_{\rm fil}$. As a consequence, galaxy clusters are the destinations for SRs, though these environments are not where they originally formed.

The small fraction of SRs with $D_{\rm fil}>2$\,Mpc are still massive, old, quenched galaxies. These SRs might be formed by environmental processes long ago, or they are currently within filaments or groups but missed by our GAMA sample.

Finally, SRs that are extremely close to filaments and nodes are affected by the local environment: at both fixed $D_{\rm node}<0.5$\,Mpc and $D_{\rm fil}<0.5$\,Mpc, the fraction of SRs increases for higher local galaxy density (Section~\ref{Fraction of SRs in environmental planes}). This result supports the role found for the local environment by previous studies in affecting galaxy spin, although this is secondary to internal galaxy properties \citep{Wang2020,vandeSande2021b,Santucci2023,Croom2024} and also secondary to the effect of filaments according to this work.

\section{Summary and conclusions}
\label{Summary and conclusions}

We study the link between galaxy angular momentum and environment, focussing on the role of large-scale structure in affecting stellar spin amplitudes ($\lambda_{\rm R_e}$) and the formation of slow rotators. We use the GAMA spectroscopic redshift survey to reconstruct the cosmic web and the SAMI Galaxy Survey to measure stellar spins from spatially-resolved stellar kinematics.

Our sample comprises 1094 SAMI galaxies with masses $\log(M_\star/M_{\odot})>9.5$, of which 976 are FRs and 118 are SRs. We investigate $\lambda_{\rm R_e}$ and the distributions of FRs and SRs as functions of galaxy internal properties (stellar age and mass), local environment metrics (local galaxy density, halo mass, central/satellite/isolated classification) and large-scale environment metrics (distance to the closest node, filament, and void in the cosmic web). The fractions of SRs are analysed within the $D_{\rm fil}$--$D_{\rm node}$ plane and local versus large-scale environmental planes. The main results are:
\begin{enumerate}[left=-3pt .. \parindent]
    \item[(i)] FRs show kinematic segregation within the cosmic web: they tend to have larger $\lambda_{\rm R_e}$ values (i.e.\ to be more rotation-dominated) going from nodes to filaments to voids. This applies to both the low-mass and high-mass FR samples.
    \item[(ii)] $D_{\rm fil}$ is the environment metric that most significantly correlates with $\lambda_{\rm R_e}$, more so than local galaxy density or distances to the closest node or void. However, stellar population age remains the dominant parameter affecting stellar spin, in agreement with previous studies.
    \item[(iii)] Two-sample K-S tests only find significant differences in the distributions of SR and $M_{\star}$-matched FR galaxies for the large-scale environment metrics; the two kinematic types are consistent with belonging to the same distribution in terms of local environment metrics. 
    \item[(iv)] In the $D_{\rm fil}$--$D_{\rm node}$ plane, the fraction of SRs increases closer to filaments and nodes. About 95\% of SRs can be found at $D_{\rm fil}\leq2$\,Mpc. 
    \item[(v)] SRs are spread throughout the $\Sigma_{\rm 5}$--$D_{\rm node}$, $M_{\rm halo}$--$D_{\rm node}$ and local type--$D_{\rm node}$ planes, with the most massive, oldest, and lowest-SFR SRs being closer to nodes and centrals in high-density regions and high-mass halos. In the $\Sigma_{\rm 5}$--$D_{\rm fil}$, $M_{\rm halo}$--$D_{\rm fil}$ and local type--$D_{\rm fil}$ planes, SRs are concentrated at $D_{\rm fil}\leq2$\,Mpc, but they cover all local galaxy densities, halo masses, and local types (similarly to FRs).
    \item[(vi)] At fixed $D_{\rm node}$ or fixed $D_{\rm fil}$, the fraction of SRs increases for higher stellar masses. At fixed $M_{\star}$, $f_{\rm SR}$ increases for smaller $D_{\rm fil}$ for massive galaxies with $\log(M_{\star}/M_{\odot})>10.7$; this trend is not significant for $D_{\rm node}$.
    \item[(vii)] At fixed $D_{\rm node}<0.5$\,Mpc or fixed $D_{\rm fil}<0.5$\,Mpc, the fraction of SRs increases for higher local galaxy densities, although the trend might be driven by stellar mass or age. At larger distances from nodes and filaments, there is no clear trend with local environment. At fixed $\Sigma_{5}$, $f_{\rm SR}$ increases for smaller $D_{\rm fil}$ values for all galaxies.
\end{enumerate}  

\noindent In conclusion, our results show how the large-scale environment affects the spin amplitude of galaxies, playing a secondary but independent role relative to the key internal galaxy properties, age and mass. $D_{\rm fil}$ is clearly the dominant environmental metric correlating with the fraction of SRs, pointing to filaments being important cosmic environments for the formation of SRs before they reach the more extreme node environments where they are commonly found. The local environment is found to impact stellar spins only for galaxies very close to filaments or nodes.

This study supports scenarios where physical processes, such as pre-processing by mergers, that mainly occur within filaments (and so correlate with $D_{\rm fil}$) initiate the formation of SRs, rather than processes that mainly occur in the more extreme node environments (traced by $D_{\rm node}$). However, at present we only have clues to SR formation scenarios due to the relatively small samples available for analyses. Upcoming large IFS galaxy surveys, such as the Hector survey \citep{Bryant2020,Bryant2024}, massive spectroscopic redshift surveys, such as the Wide Area VISTA Extra-galactic Survey (WAVES; \citet{Driver2016}) and the 4MOST Hemisphere Survey of the nearby Universe (4HS; \citet{Taylor2023}), and future facilities, such as the proposed Wide-field Spectroscopic Telescope \citep{mainieri2024widefield}, will be able to draw stronger conclusions on the relation between galaxy angular momentum and the large-scale structures of our Universe.

\section*{Acknowledgements}
We thank the referee for the constructive report. This research was supported by the Australian Research Council Centre of Excellence for All Sky Astrophysics in 3 Dimensions (ASTRO~3D, CE170100013). The SAMI Galaxy Survey is based on observations made at the Anglo-Australian Telescope. The Sydney-AAO Multi-object Integral field spectrograph (SAMI) was developed jointly by the University of Sydney and the Australian Astronomical Observatory, and funded by ARC grants FF0776384 (Bland-Hawthorn) and LE130100198. The SAMI input catalogue is based on data taken from the Sloan Digital Sky Survey, the GAMA Survey, and the VST/ATLAS Survey. The SAMI Galaxy Survey website is http://sami-survey.org/. This study uses data provided by AAO Data Central (http://datacentral.org.au/) and the \href{http://www.python.org}{Python} programming language \citep{vanrossum1995}. We acknowledge the use of {\sc \href{https://pypi.org/project/numpy/}{numpy}} \citep{harris+2020}, {\sc \href{https://pypi.org/project/scipy/}{scipy}} \citep{jones+2001}, {\sc \href{https://pypi.org/project/matplotlib/}{matplotlib}} \citep{hunter2007}, {\sc \href{https://pypi.org/project/astropy/}{astropy}} \citep{astropyco+2013}, {\sc \href{https://pingouin-stats.org/}{pingouin}} \citep{Vallat2018}, {\sc \href{https://scikit-learn.org/stable/}{scikit-learn}} \citep{Pedregosa2012} and {\sc \href{http://www.star.bris.ac.uk/~mbt/topcat/}{topcat}} \citep{taylor2005}. SB acknowledges the support from the Physics Foundation through the Messel Research Fellowship. SO acknowledges support from the NRF grant funded by the Korea government (MSIT) (No. RS-2023-00214057 and No. RS-2025-00514475).
JJB acknowledges support of an Australian Research Council Future Fellowship (FT180100231). CW was supported in part by the National Science Foundation under Grant No. NSF PHY-1748958. JvdS acknowledges support from an Australian Research Council Discovery Early Career Research Award (DE200100461) funded by the Australian Government.

\section*{Data availability}

The SAMI data used in this paper are publicly available at \href{https://docs.datacentral.org.au/sami}{SAMI Data Release 3} \citep{Croom2021}. Ancillary data are from \href{http://gama-survey.org}{GAMA Data Releases 3 \& 4} \citep{Baldry2018,Driver2022}.  

\section*{Author Contribution Statement}

SB devised the project, carried out the analysis and drafted the paper. SB, SMC and MC contributed to data analyses and interpretation of the results. JJB, SMC, NL and JvdS provided key support to all the activities of the SAMI Galaxy Survey (`builder status'). All authors discussed the results and commented on the manuscript.


\bibliographystyle{mnras}
\bibliography{biblioSAMI} 


\appendix

\section{Comparison of local and large-scale environment metrics}
\label{Local environments within cosmic regions}

\begin{figure}
\centering
\includegraphics[width=1.13\columnwidth]{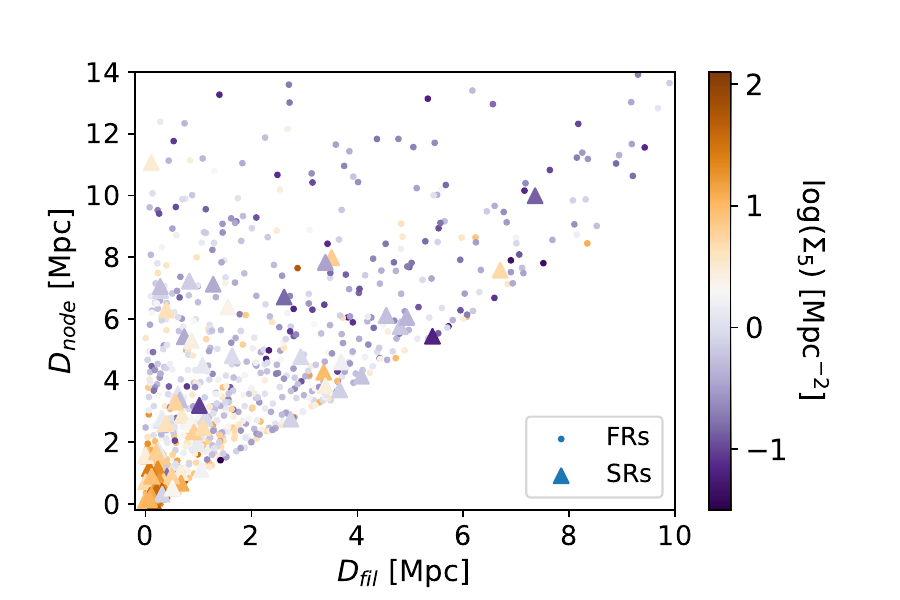}
\includegraphics[width=1.13\columnwidth]{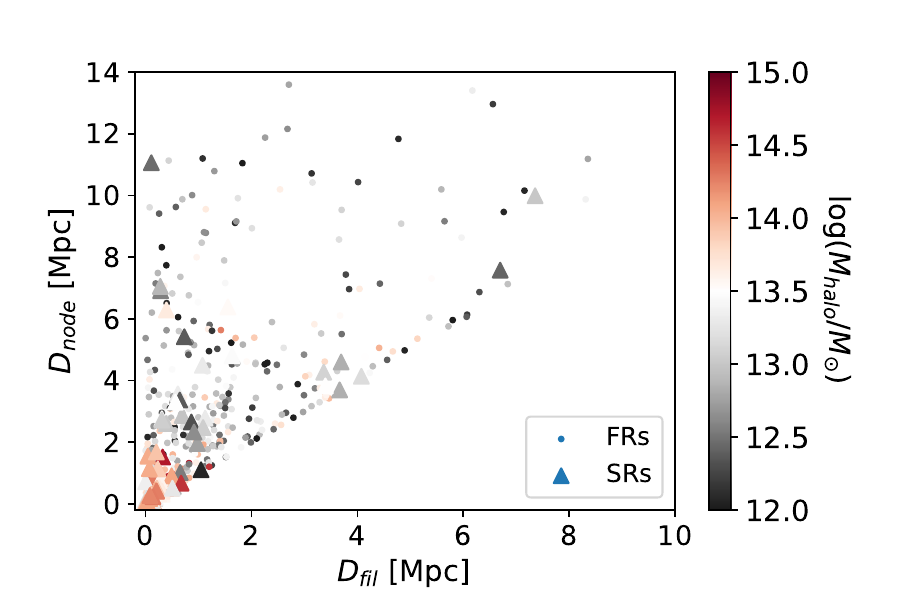}
\includegraphics[width=1.13\columnwidth]{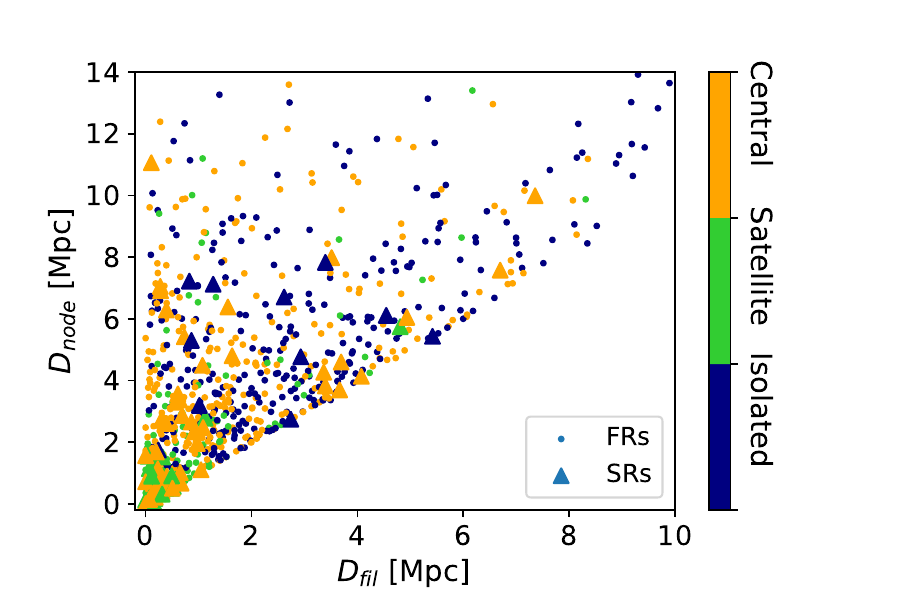}
\caption{$D_{\rm fil}$--$D_{\rm node}$ plane colour-coded by local galaxy density (top panel), halo mass (middle panel), and local type (bottom panel). FRs are shown as dots and SRs are shown as triangles.}
\label{DfilDnode_plane_local}
\end{figure}

Since our goal is to explore SRs in the context of local versus large-scale environments, we check the values of the local environment metrics as functions of galaxies' large-scale environment metrics. Figure~\ref{DfilDnode_plane_local} shows the large-scale environment $D_{\rm fil}$--$D_{\rm node}$ plane colour-coded by local metrics: the top panel shows local galaxy density; the middle panel shows halo mass; and the bottom panel shows local type. Galaxies close to nodes and filaments generally belong to high-density local environments, are in massive halos, and tend to be centrals, as we expect from the hierarchical formation of structure in the Universe. 

\section{Kinematic residuals}
\label{Kinematic residuals}
\begin{figure}
\includegraphics[scale=0.30]{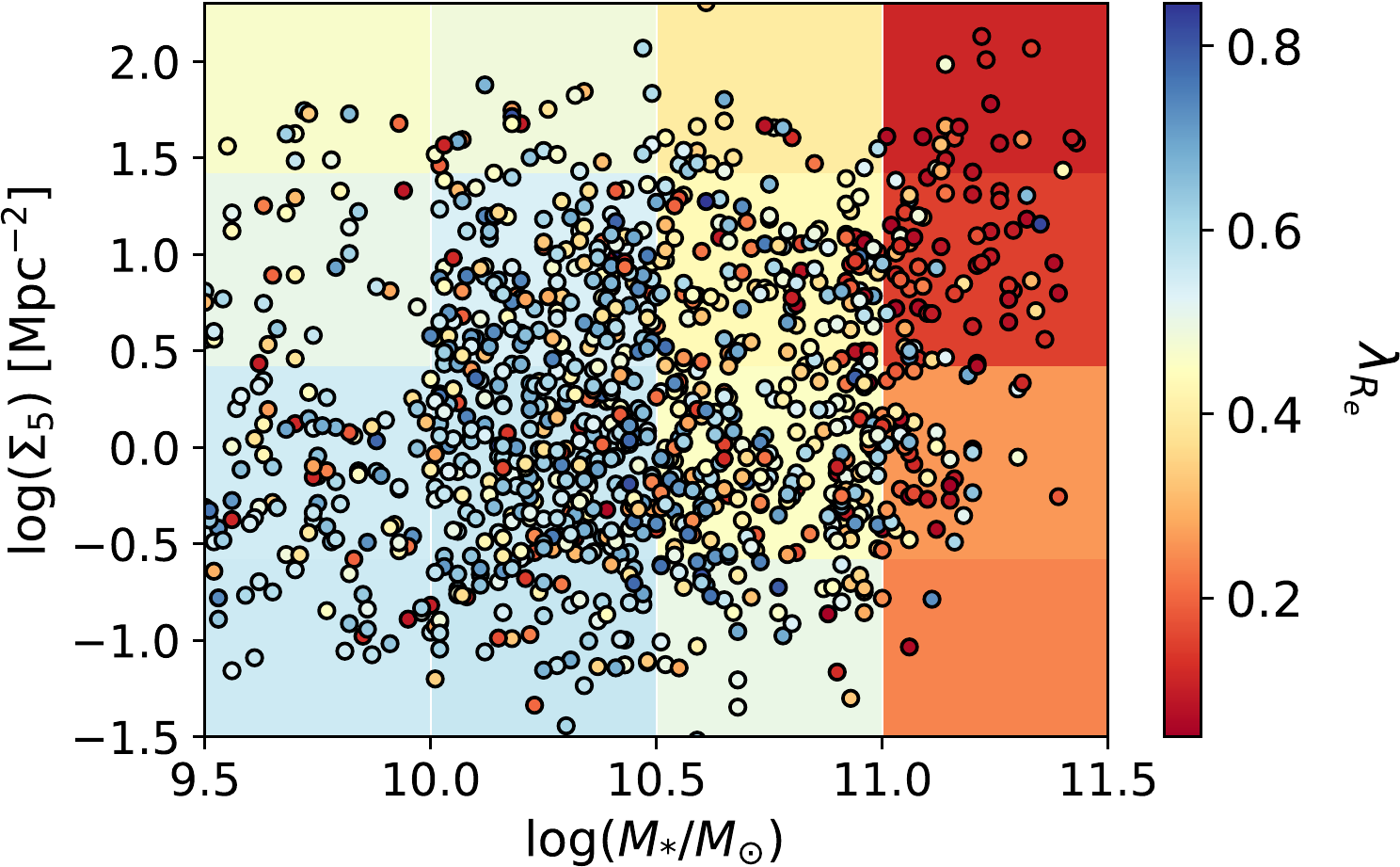}

\vspace{0.5cm}

\includegraphics[scale=0.30]{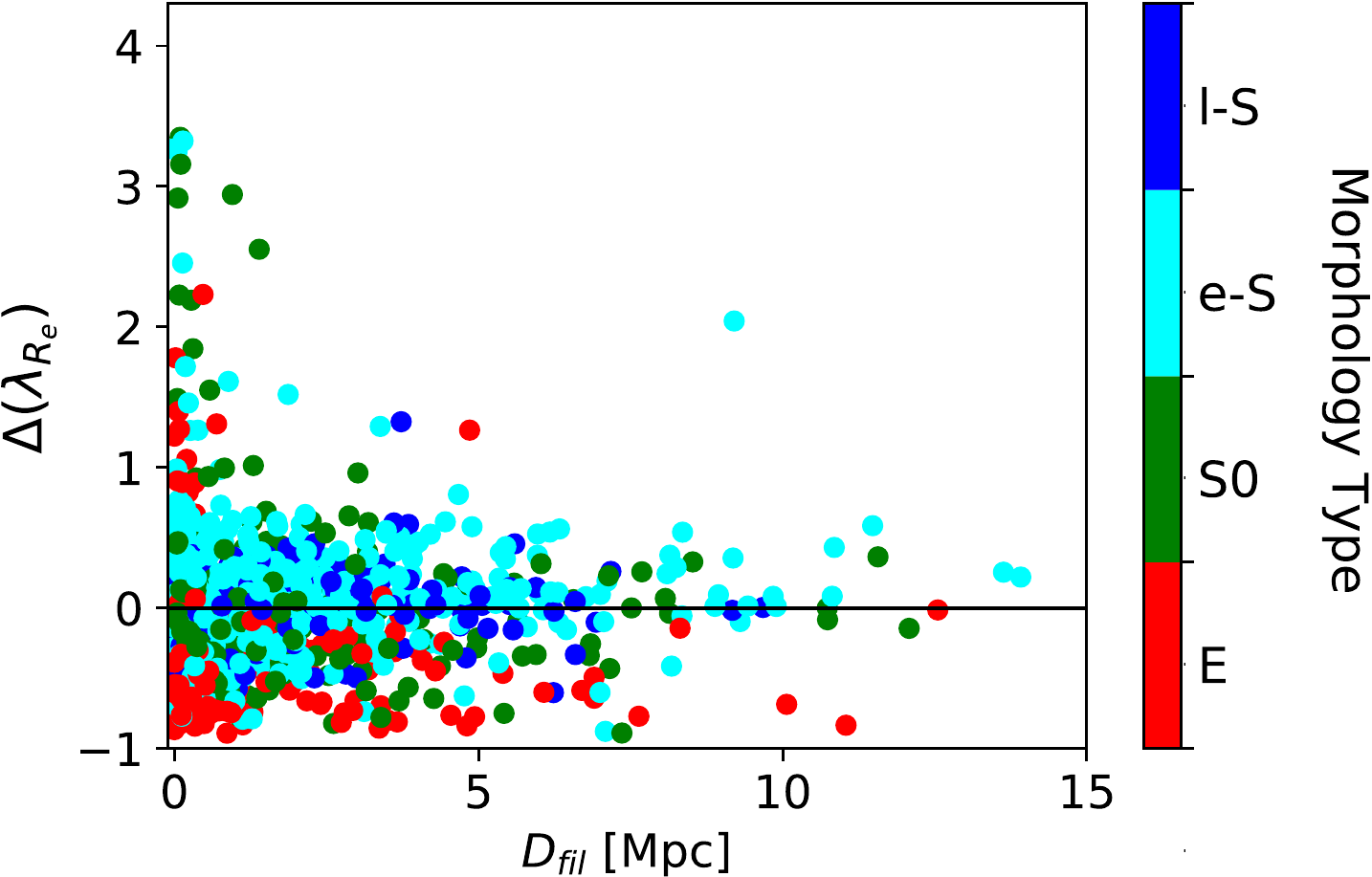}
\caption{Top panel: $\lambda_{\rm R_e}$ distribution in the $M_{\star}$ versus $\Sigma_5$ plot where the medians of each bin are shown as the binned colour map underlying the individual galaxies (black circled dots). Bottom panel: kinematic residuals $\Delta({\lambda_{\rm R_e}})$ as a function of $D_{\rm fil}$ colour-coded by morphological type, showing that the largest residuals from zero (black line) are found for smaller $D_{\rm fil}$ distances and for spirals and S0 galaxies.}
\label{DeltaKin}
\end{figure}

As a complementary analysis to the partial correlation coefficients shown in Section~\ref{Correlations}, we study the residual correlations of $D_{\rm fil}$ with stellar mass and local galaxy density, defining the kinematic residuals:
\begin{equation}
\Delta({\lambda_{\rm R_e},i})=\frac{\lambda_{\rm R_e},_i-\langle \lambda_{\rm R_e}\rangle_{\rm bin}}{\langle\lambda_{\rm R_e}\rangle_{\rm bin}}
\end{equation}
where $\langle\lambda_{\rm R_e}\rangle_{\rm bin}$ is the median over a stellar mass and local galaxy density bin. The top panel of Figure~\ref{DeltaKin} shows the $\lambda_{\rm R_e}$ distribution in the $M_{\star}$ versus $\Sigma_5$ plot where the medians of each bin are shown as the binned colour map underlying the individual galaxies. The bottom panel of Figure~\ref{DeltaKin} shows 
$\Delta({\lambda_{\rm R_e}})$ as a function of $D_{\rm fil}$ colour-coded by morphological type: the largest residuals from zero are positive, meaning  higher spin than the median for their bin, and are found for smaller distances of $D_{\rm fil}$ for spirals and S0 galaxies. These same conclusions are found if we kernel-smooth the distribution of $\lambda_{\rm R_e}$ over stellar mass and local galaxy density instead of using the medians in the $M_{\star}-\Sigma_5$ bins. This analyis is in agreement with the significant correlation of residuals only for $D_{\rm fil}$ ($p$-value $=0.009$) while controlling for $M_\star$+$\Sigma_{5}$ from Figure~\ref{PartialCorrelations} of Section~\ref{Correlations}.



\bsp	
\label{lastpage}
\end{document}